\documentclass[times,twocolumn]{aastex61}
\newcommand{\kms}{\mbox{km\thinspace s$^{-1}$}}     

\newcommand{\msun}{\mbox{$M_\odot$}}

\newcommand{\hst}{\emph{HST}}
\newcommand{\chandra}{\emph{Chandra}}

\newcommand{\ergs}{\mbox{${\rm erg}~{\rm s}^{-1}$}}

\newcommand{\B}{\mbox{$B_{435}$}}
\newcommand{\R}{\mbox{$R_{625}$}}
\newcommand{\br}{\mbox{$\B\!-\!\R$}}
\newcommand{\ha}{\mbox{H$\alpha$}}
\newcommand{\hr}{\mbox{$\ha\!-\!\R$}}

\newcommand{\note}[1]{\marginpar{\small\em $\Leftarrow$ \raggedright #1}}
\renewcommand{\note}[1]{}
\newcommand{\ignore}[1]{}

\shortauthors{Lugger et al.}
\shorttitle{Faint Chandra Sources in NGC~6752}

\begin{document}

\published{in the Astrophysical Journal, 2017, 241, 53}

\title{Identification of Faint \emph{Chandra} X-ray Sources in the Core-Collapsed Globular Cluster NGC~6752}
\author{Phyllis M. Lugger}
\affiliation{Department of Astronomy, Indiana University, 727 E. Third St.,
Bloomington, IN 47405, USA; lugger@indiana.edu}

\author{Haldan N. Cohn} 
\affiliation{Department of Astronomy, Indiana University, 727 E. Third St.,
Bloomington, IN 47405, USA; cohn@indiana.edu}

\author{Adrienne M. Cool}
\affiliation{Department of Physics and Astronomy, San Francisco State
University, 1600 Holloway Avenue, San Francisco, CA 94132, USA; cool@sfsu.edu}

\author{Craig O. Heinke}
\affiliation{Department of Physics, University of Alberta, Edmonton, AB T6G
2G7, Canada; heinke@ualberta.ca}

\author{Jay Anderson}
\affiliation{Space Telescope Science Institute, 3700 San Martin Dr.,
  Baltimore, MD 21218, USA; jayander@stsci.edu}

\begin{abstract}

We have searched for optical identifications for 39 \chandra\ X-ray
sources that lie within the 1\farcm9 half-mass radius of the nearby
($d = 4.0~{\rm kpc}$), core-collapsed globular cluster, NGC~6752,
using deep \emph{Hubble Space Telescope} ACS/WFC imaging in \B, \R,
and \ha.  Photometry of these images allows us to classify candidate
counterparts based primarily on color-magnitude and color-color
diagram location.  The color-color diagram is particularly useful for
quantifying the \ha-line equivalent width. In addition to recovering
11 previously detected optical counterparts, we propose 20 new optical
IDs. In total, there are 16 likely or less certain cataclysmic
variables (CVs), nine likely or less certain chromospherically active
binaries, three galaxies, and three active galactic nuclei (AGNs). The
latter three sources, which had been identified as likely CVs by
previous investigations, now appear to be extragalactic objects based
on their proper motions. As we previously found for NGC~6397, the CV
candidates in NGC~6752 fall into a bright group that is centrally
concentrated relative to the turnoff-mass stars and a faint group that
has a spatial distribution that is more similar to that of the
turnoff-mass stars.  This is consistent with an evolutionary scenario
in which CVs are produced by dynamical interactions near the cluster
center and diffuse to larger radius orbits as they age.   
\end{abstract}  


\section{Introduction}

Joint \emph{Chandra X-ray Observatory} and \emph{Hubble Space
  Telescope} (\hst) observations of globular clusters have revealed
large populations of faint X-ray sources ($L_X \lesssim
10^{33.5}~\ergs$) which include quiescent low-mass X-ray binaries
(qLMXBs), cataclysmic variables (CVs), chromospherically active
binaries (ABs), and millisecond pulsars (MSPs)
\citep{Verbunt06,Heinke10}.  These populations range in size from tens
to hundreds of objects per cluster.  The presence of these objects is
closely related to the cluster dynamics, as demonstrated by
\citet{Pooley03}, who found the existence of a strong correlation
between the source population size and the encounter rate in the
cluster core, $\Gamma \propto {\rho_0}^2 {r_c}^3 {v_0}^{-1}$, where
$\rho_0$ is the central density, $r_c$ is the core radius, and $v_0$
is the central velocity dispersion.  \citet{Pooley06} subsequently
found evidence that the majority of CVs in dense globular clusters
were produced dynamically.  Clusters that undergo core collapse pass
through phases of extremely high central density and are thus expected
to undergo repeated bursts of close binary production in their cores.
It is therefore of great interest to examine the faint X-ray source
populations in core-collapsed clusters.

In a previous study, we carried out a deep \hst\ ACS/WFC imaging study
of the nearest core-collapsed globular cluster NGC~6397 in the filters
F435W (\B), F625W (\R), and F658N (\ha), in which we identified
optical counterparts to 69 of the 79 \chandra\ sources that lie within the
half-mass radius \citep{Cohn10}.  A striking finding of this study was
that there is a bimodal distribution of CVs, consisting of a brighter
group of six for which the optical emission is dominated by
contributions from the secondary and the accretion disk, and a fainter
group of seven for which the white dwarf dominates the optical
emission.  The brighter CVs are much more concentrated towards the
cluster center than the fainter CVs.  We speculated that this may
be the result of dynamical evolution in which CVs are formed in
and near the cluster core and subsequently migrate to larger distances
from the cluster center as they age and become fainter.  The faintest
CVs that we identified in our study of NGC~6397 have absolute
magnitudes around $M_R \sim 11.2$ that likely put them near the
minimum of the CV period distribution found in the Sloan Digital Sky
Survey \citep{Gaensicke09}.  \citet{Cool13} used \hst\ imaging to
search for optical counterparts to \chandra\ sources in the massive
globular cluster $\omega$\,Cen.  They reported finding 27 candidate CVs,
of which 14 lie in the magnitude range $M_R \sim 10.4 - 12.6$.  Thus,
faint CV populations that likely extend to the minimum of the CV
period distribution are known to exist in two globular clusters.

To extend the search for faint CV populations, we have investigated
NGC~6752, the second closest core-collapsed cluster at a distance of
4.0~kpc \citep{Harris96}.  The puzzling dynamical status of NGC~6752
has been addressed in a number of studies.  \citet{Djorgovski86}
classified the cluster as core-collapsed based on its ground-based
$B$-band surface-brightness profile, which shows a power-law region
outside of a resolved core.  In \citet{Lugger95}, we reexamined the
profile using ground-based $U$-band data, which is less affected by
bright giants, and found that it could be fitted by either a modified
power law (power-law + core) or a standard King model, with the former
providing a somewhat better fit than the latter.  We nevertheless
advanced the conservative interpretation that NGC~6752 is not required
to be in a post-collapse state.  \citet{Ferraro03a} used a combination
of \hst\ WFPC2 and ground-based images to construct surface-brightness
and surface-density profiles, finding that these are best fitted by a
double King model, viz.\ one King model to describe the inner profile
and another one to describe the outer profile.  They took this as
evidence that NGC~6752 is experiencing a post-core-collapse bounce.
Most recently, \citet{Thomson12} obtained surface-density profiles
from \hst\ ACS imaging, again finding that the profile is not well
described by a single King model.  They present a double-King-model
fit and conclude that the cluster is either in a core-collapsed or
post-core-collapse state.  They note that the double King model is
purely phenomenological, without a physical basis.  Based on this
previous work, we have undertaken a reexamination of the
surface-density profile of NGC~6752 in the present study.

The X-ray source population of NGC~6752 has been previously studied by
\citet{Pooley02}, \citet{Kaluzny09}, and \citet{Thomson12}.
\citet{Pooley02} identified a total of 19 \chandra\ sources within the
115\arcsec\ half-mass radius of the cluster to a limiting X-ray
luminosity of $L_X \sim 10^{30} ~\ergs$.  They proposed 12 optical
identifications based on archival \hst\ WFPC2 imaging and one radio
identification.  They found that 10 of the sources are likely to be
CVs, one to three are likely to be ABs, and one or two are possible
background objects.  \citet{Kaluzny09} identified a periodic variable
that they suggested as the optical counterpart to the \citet{Pooley02}
source CX19.\@ \citet{Thomson12} reanalyzed the identifications of the
\citet{Pooley02} sources using multi-wavelength imaging (FUV to
$I$-band) from the \hst\ STIS, ACS, and WFC3\@.  They identified
dwarf nova outbursts from two previously identified CVs, suggested
optical counterparts to CX8 and CX12 (and an alternate optical
counterpart to CX16), and failed to confirm suggested counterparts to
CX11 \citep{Pooley02} and CX19 \citep{Kaluzny09}.

\begin{figure*}
\epsscale{0.8}
\plotone{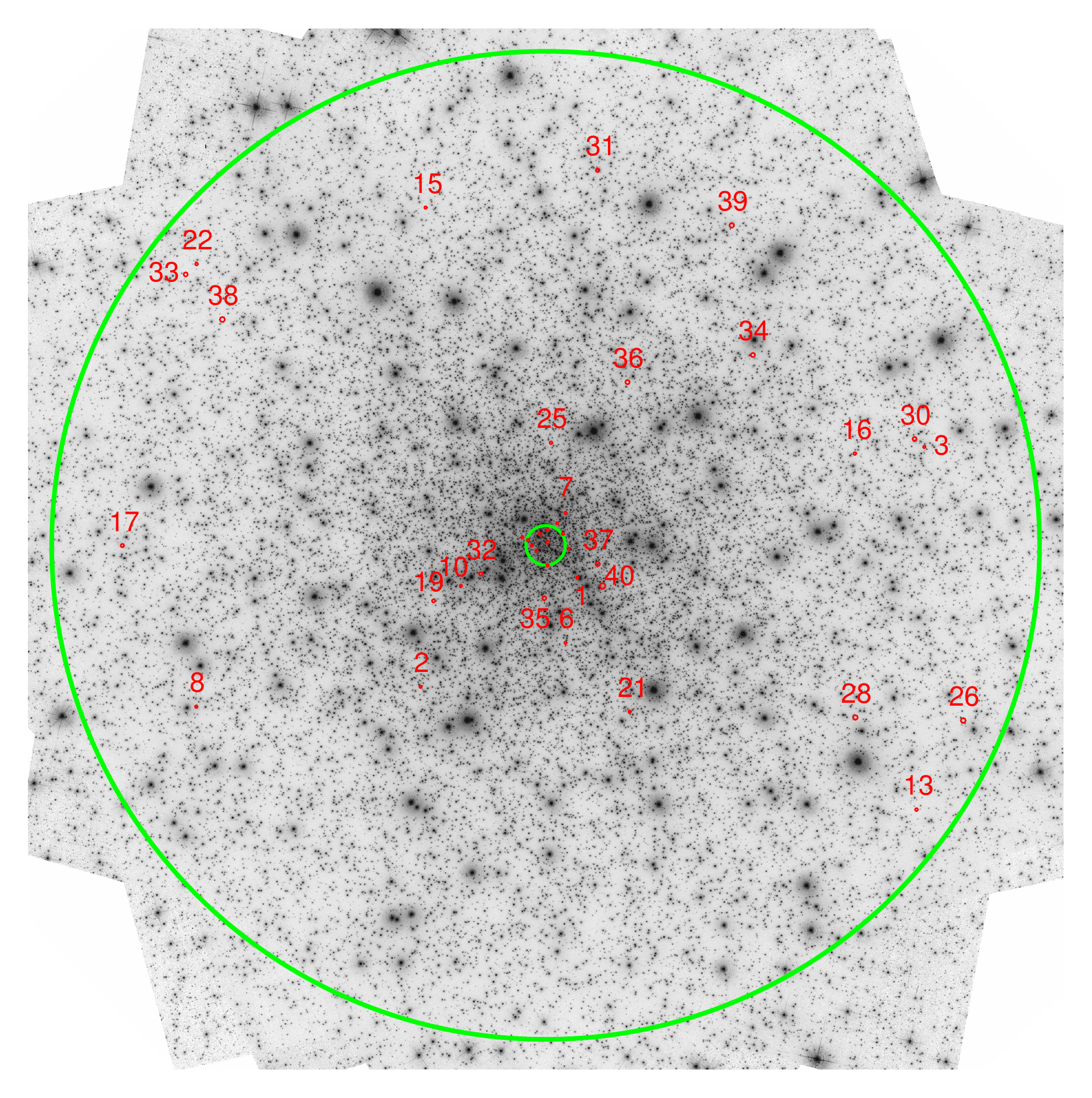}
\figcaption{Drizzle-combined \hst\ ACS/WFC \ha\ mosaic of NGC~6752
  with search regions for \chandra\ sources within the half-mass
  radius. North is up and east is to the left. The inner green circle
  represents the core radius of 4\farcs6 and the outer green circle
  represents the half-mass radius of 1\farcm9.  Source labels are
  omitted near the core for clarity.  There are a total of 39 sources
  detected within the half-mass radius.  The search region radius is
  defined as the maximum of 2.5 times the formal error circle radius
  and 0\farcs3. 
\label{f:mosaic}}
\end{figure*}

\begin{figure*}
\epsscale{0.8}
\plotone{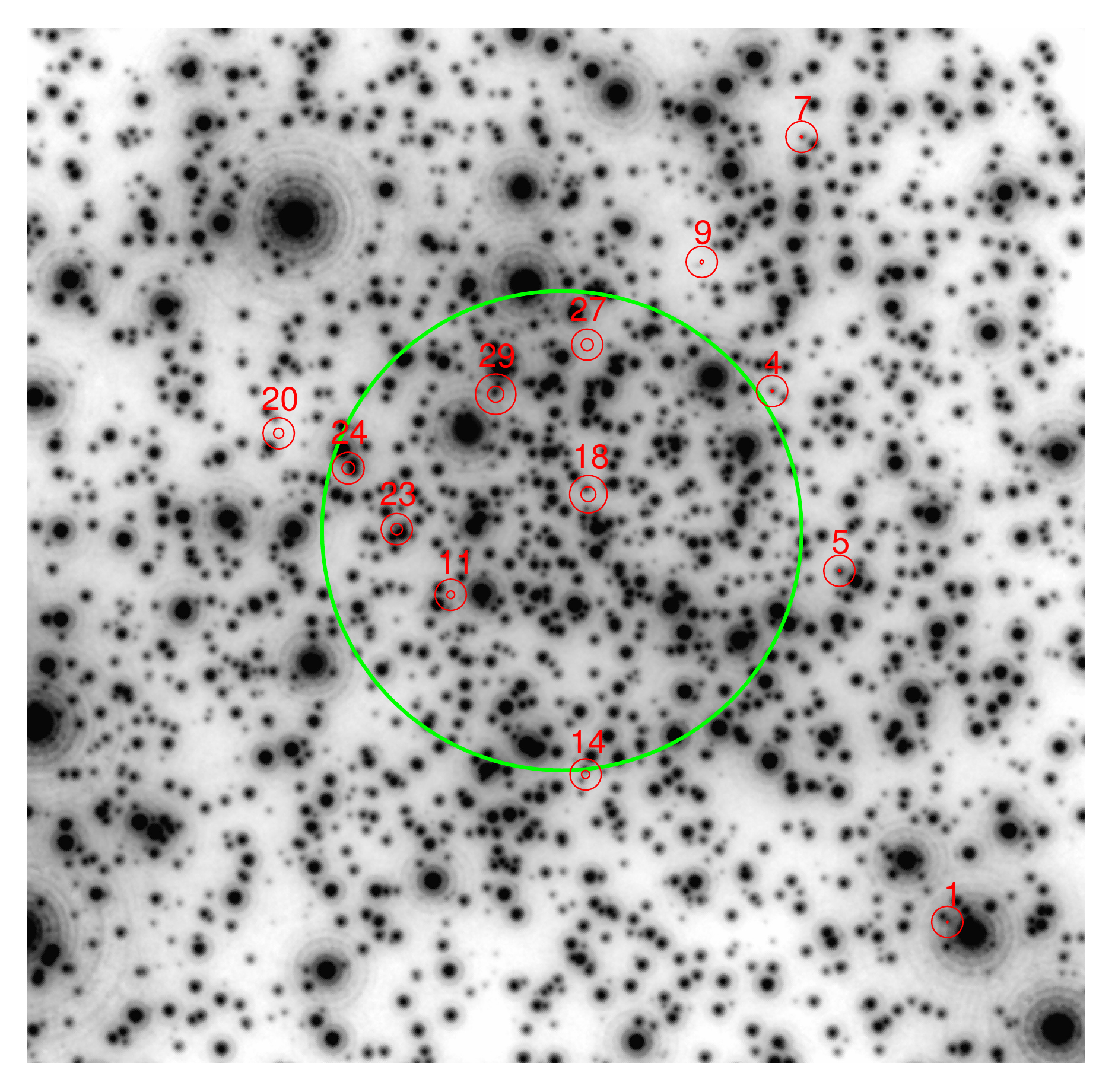}
\figcaption{The core-radius region of the drizzle-combined \ha\ mosaic
  of NGC~6752 with error circles and search regions for
  \chandra\ sources. The green circle represents the core radius of
  4\farcs6.  
\label{f:mosaic_zoom}}
\end{figure*}

The \chandra\ study by \citet{Pooley02} is based on 29\,ks of the
total of 67\,ks of ACIS exposure that is available for this cluster.
\citet{Forestell14} have analyzed the complete dataset, detecting 39
sources within the half-mass radius, to a limiting luminosity of $L_X
\approx 5 \times 10^{29} ~\ergs$.  In order to search for counterparts
of this deeper set of \chandra\ sources, we have made use of the
ACS/WFC imaging database that is also being used to search for
helium-core white dwarfs in NGC~6752 \citep{Hernandez13}, results of
which will be reported elsewhere. 

In the following sections, we describe the \chandra\ and \hst\ datasets
that we use in this study, the method that we use to analyze the
\hst\ dataset to find optical counterparts to the \chandra\ sources,
the set of counterparts that results from this analysis, the optical
variability of the counterparts, and the spatial distribution of the
stars and X-ray sources in NGC~6752.

\section{Data \label{data}}

The \chandra\ imaging used in this study is the combination of
Observation IDs 948 (PI: Lewin) and 6612 (PI: Heinke), which were
carried out with the ACIS-S instrument.  The exposure times for these
datasets are 29\,ks and 38\,ks respectively.  The processing of these
data is described in detail by \citet{Forestell14}.  Source detections
were made with the {\tt wavdetect} and {\tt pwdetect} software
utilities.  This resulted in a catalog of 39 sources within the
half-mass radius of NGC 6752, extending the previous catalog
numbering by \citet{Pooley02} in order of decreasing source
brightness.\footnote{Note that CX12 from \citet{Pooley02} is divided
  into three sources in this new catalog, viz.\ CX20, CX23, and CX24.}
We restrict the present study to the analysis of these 39 sources.

The optical imaging used in this study is the \hst\ GO-12254 dataset
(PI: Cool), which provides deep, highly dithered ACS/WFC imaging of
the half-mass radius region of NGC~6752 in \B, \R, and \ha.  The
dither strategy was designed to recover the full angular resolution of
the \hst, which is advantageous for performing photometry in the very
crowded central regions of NGC~6752.  The images were obtained over
six 2-orbit visits, spread over 180\,d, in order to sample the stellar
point-spread function (PSF) at a range of roll angles.  When combined
by a drizzling technique, the resulting mosaic images are free of
diffraction spikes and similar PSF artifacts.  The dataset consists of
6 short \B\ (10\,s), 12 long \B\ (380\,s), 6 short \R\ (10\,s), 12
long \R\ (360\,s), and 24 \ha\ (12 each of 724\,s and 820\,s)
exposures.  The short exposures were designed to provide accurate
photometry for stars above the main-sequence turnoff (MSTO)\@.  With
the large number of \ha\ frames, the PSF sampling is particularly
complete for this band.

\section{Analysis Method \label{analysis}}

\subsection{Photometry \label{photometry}}

The \hst\ data were analyzed using software based on the program
developed for the ACS Globular Cluster Treasury project, described in
\citet{Anderson08}; we have previously used this software in the
search for counterparts to \chandra\ sources in NGC~6397.  The
reductions were done using an updated version of this software known
as KS2\@. Since our best coverage was in \ha, we did the first several
iterations of star finding on those images.  In order to capture very
faint stars on both the main sequence and the white dwarf sequence, 
we followed this with an iteration of star finding using the \R\
exposures alone and then a final iteration using the \B\ and \R\
exposures combined. A total of 68,439 stars were detected in a
mosaic that covers the half-mass radius (115\arcsec) and somewhat
beyond, reaching to 150\arcsec\ in the corners.  

KS2 uses multiple methods for measuring magnitudes.  For stars that
are well exposed in individual images, we adopted KS2 photometry
derived from fitting the PSF to stars in the individual images and
averaging the results using sigma clipping to remove outlying values
due to cosmic rays, defective pixels, etc.  For faint stars, we
adopted the KS2 photometry derived from a simultaneous fit of the PSF
to all exposures (see \citealt{Anderson08} for details).  In the
\B\ and \R\ bands, photometry was performed separately for the short
and long frames.

\begin{figure*}
\centering
\includegraphics*[clip,viewport=18 144 592 718,width=5.5in]{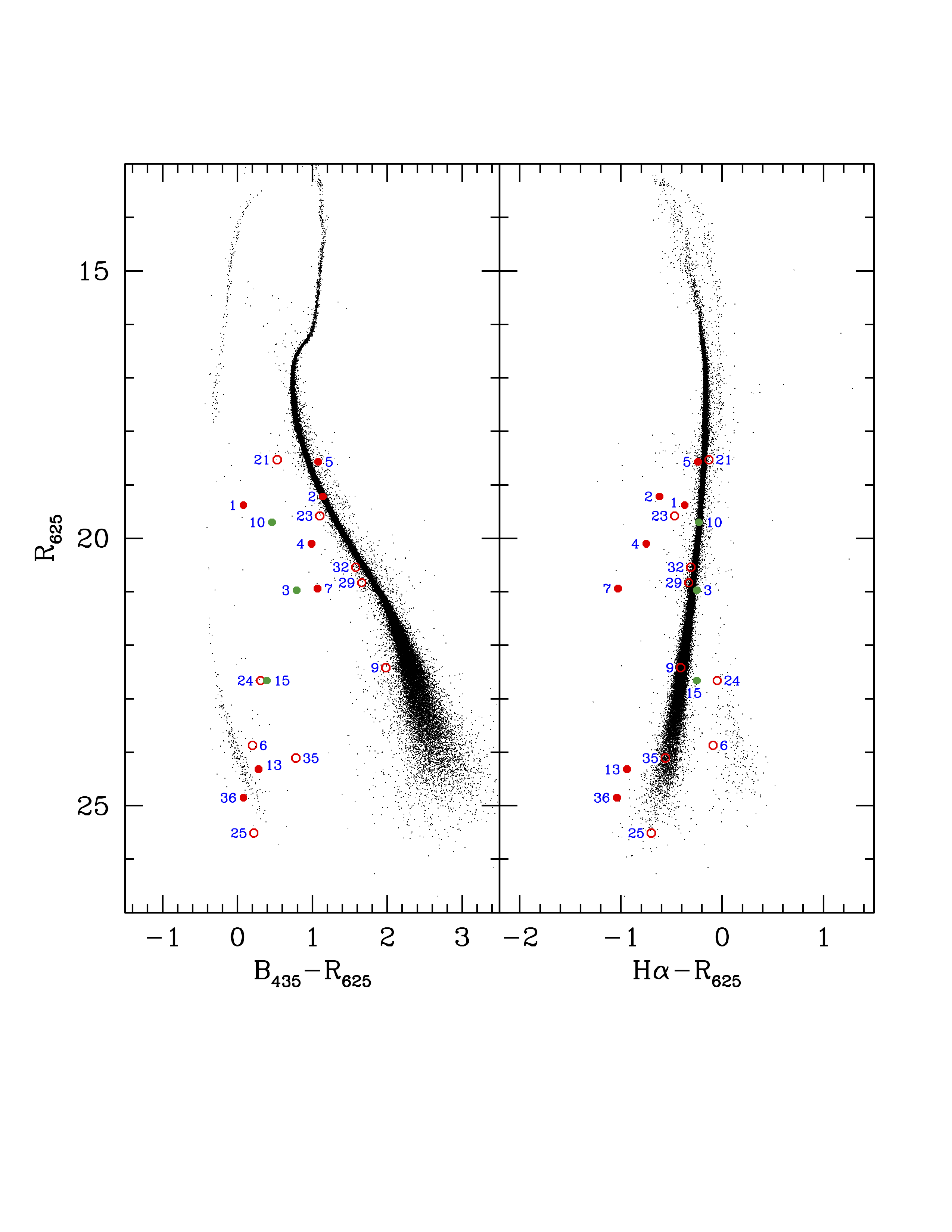}
\figcaption{Proper-motion cleaned color-magnitude diagrams for stars
  within the half-mass radius of NGC~6752 and CV identifications.  The
  candidates have been selected based on their blue color and/or
  \ha\ excess.  Open symbols indicate less certain CV identifications,
  either due to a weak or absent \ha\ excess and/or uncertain
  photometry.  Note that in the right panel, the bright CVs mostly lie
  to the \ha-excess side of the MS, while the faint CVs mostly lie to
  the \ha-excess side of the WD clump, which itself lies to the
  \ha-deficit side of the MS. All candidate counterparts are shown,
  independent of their proper-motion status. The counterparts to CX3,
  CX10, and CX15 have proper motions that are consistent with the
  extragalactic frame. The counterpart to CX29 has a proper motion
  that is not consistent with either the cluster frame nor the
  extragalactic frame. Several other counterparts have undetermined
  proper motions (see Table~\ref{t:counterparts}).
\label{f:CMD_CV}}
\end{figure*}

\begin{figure*}
\centering
\includegraphics*[clip,viewport=18 144 592 718,width=5.5in]{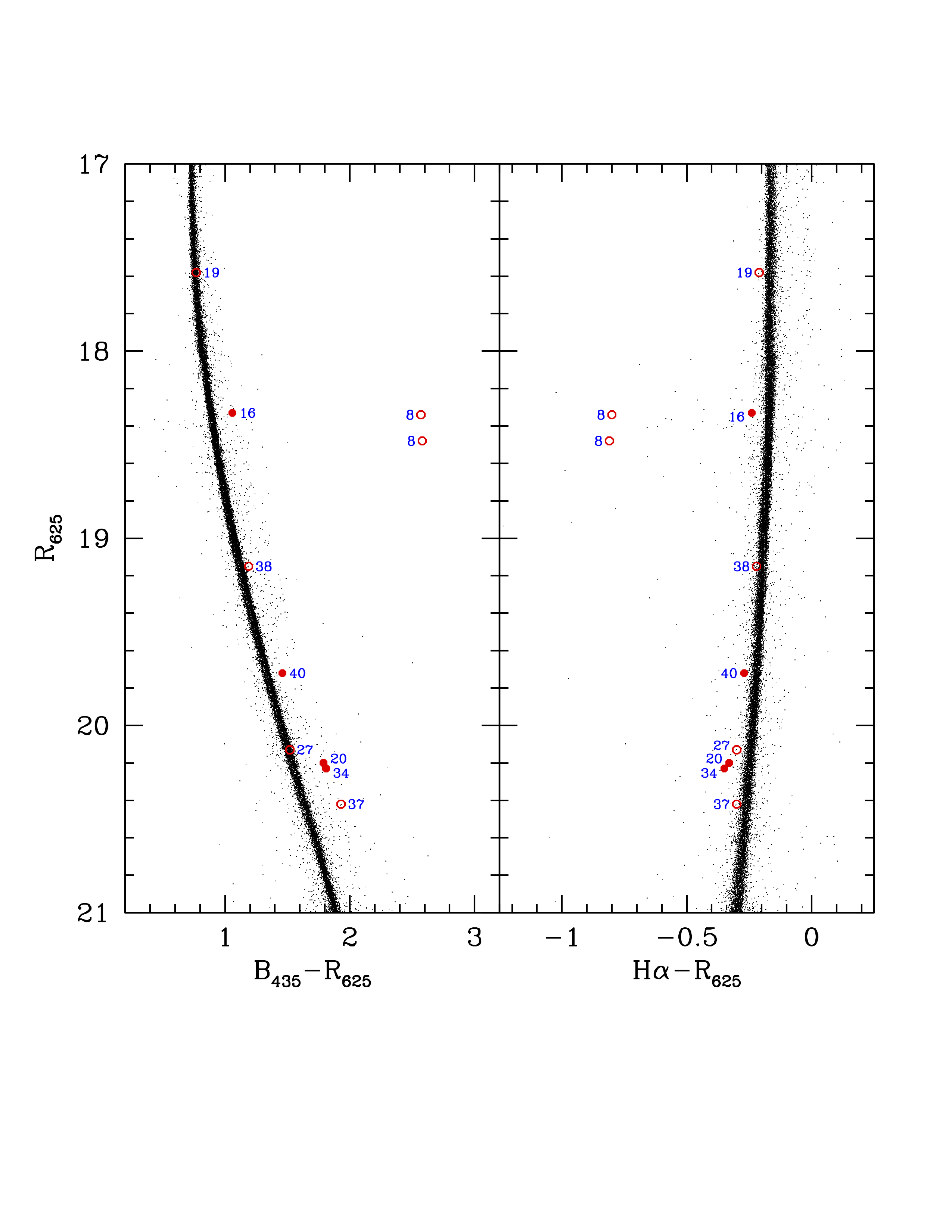}
\figcaption{Proper-motion cleaned color-magnitude diagrams for stars
  within the half-mass radius of NGC~6752 and AB identifications.  The
  candidates have been selected based on their red color and generally
  small \ha\ excess. All candidate counterparts are shown, independent
  of their proper motion status. The two counterparts to CX8 have an
  apparent proper motion that is inconsistent with the cluster mean. 
\label{f:CMD_AB}}
\end{figure*}

Photometric calibration to the {\small VEGAMAG} system was performed
by doing aperture photometry on moderately bright, isolated stars
within a 0\farcs1 radius aperture, finding the aperture correction to
an infinite radius aperture from \citet{Sirianni05}, calculating the
median offset between the KS2 photometry and the aperture photometry,
and applying the calibrations from the \hst\ ACS calibration
website.\footnote{http://www.stsci.edu/hst/acs/analysis/zeropoints} We
produced drizzle-combined mosaic images using the STScI PyRaF routine
{\tt astrodrizzle} from the {\tt drizzlepac} package.  The
drizzle-combined images were oversampled by a factor of two, in order
to increase the effective resolution.  The resulting supersampled
mosaics have an approximately 12,000$\times$12,000 format and cover an
approximately circular field of diameter 5\arcmin\ with a pixel scale
of 0\farcs025.  Figures \ref{f:mosaic} and \ref{f:mosaic_zoom} show
the drizzle-combined mosaic of 24 \ha\ frames, together with error
circles for the 39 \chandra\ sources within the half-mass radius.

Color-magnitude diagrams (CMDs) were constructed 
using the \R\ magnitudes, and the \br\ and \hr\ color indices.  
For faint stars ($\R > 21$), we used the photometry derived from
simultaneous fits to all of the long exposures. For
intermediate-brightness stars ($18 < \R < 21$), we used photometry
derived from measurements in individual long exposures. For brighter
stars ($\R < 18$), most of which are saturated in the long exposures,
we used photometry derived from individual short exposures. The very
brightest stars ($\R \lesssim 14.2$) are saturated even in short
exposures, and are thus less well measured. 

Figures \ref{f:CMD_CV} and \ref{f:CMD_AB} show the
resulting CMDs, where proper-motion cleaning has been applied to
filter out field stars (see \S\ref{astrometry}).  Without
proper-motion cleaning, the (\br, \R) CMD reaches deepest for the
bluest stars, since the faintest red main-sequence (MS) stars are
below the detection limit in \B.  The upper part of the white dwarf
(WD) cooling sequence is clearly detected in the (\br, \R) CMD,
extending to nearly 10 mag below the MSTO in \R.  There is also a
suggestion of a second WD sequence above the primary
carbon-oxygen-core WD sequence, which \citet{Hernandez13} interpreted as
a helium-core WD sequence.

\begin{figure*}
\centering
\includegraphics*[clip,viewport=18 144 592 718,width=5.5in]{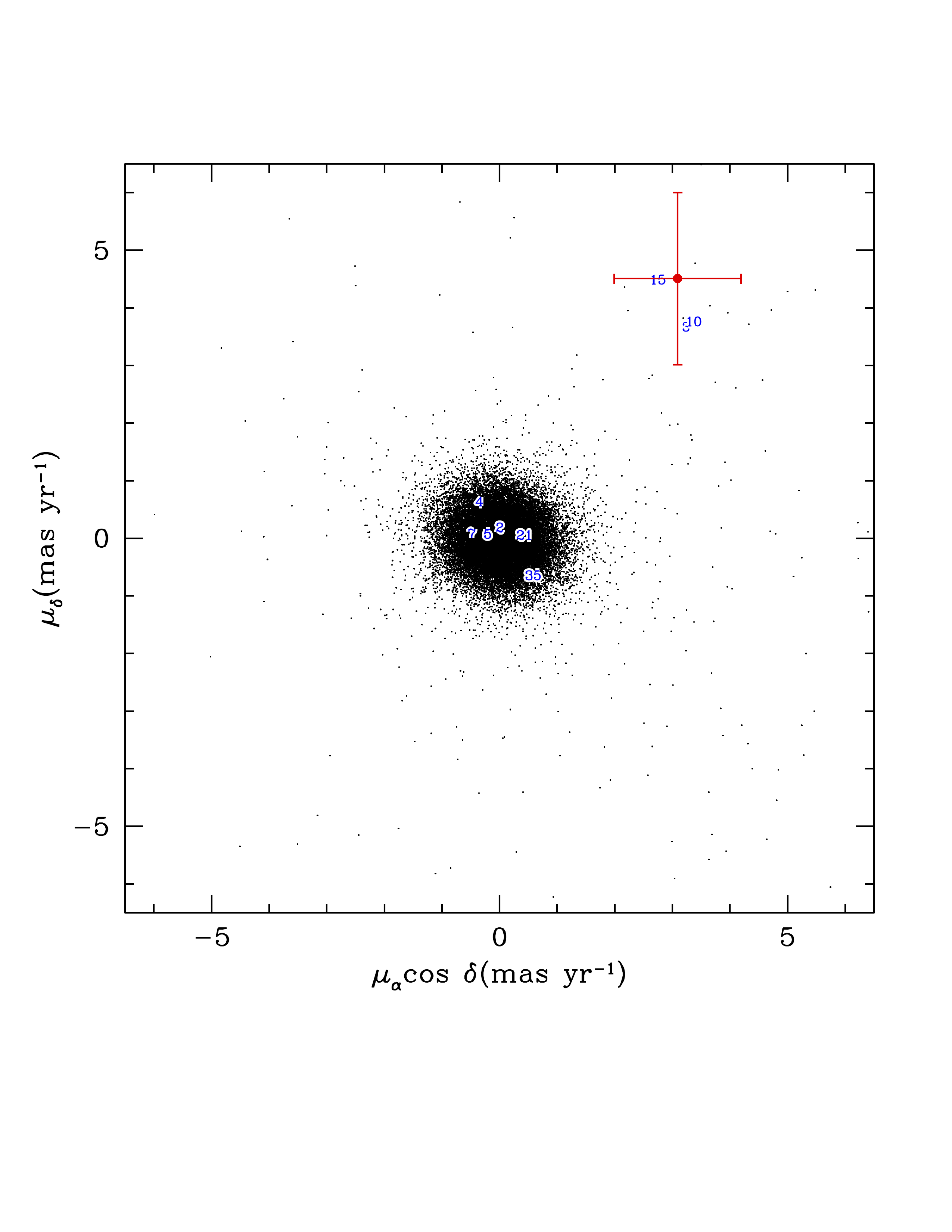}
\figcaption{Proper-motion component distribution for stars in the
  magnitude range $18 \le \R < 23$.  The zero-point corresponds to the
  systemic cluster motion. The red point and error bars indicate the
  mean and standard error of the mean of the measured proper motions
  of eight elliptical galaxies, six of which are in the mosaic field
  and two of which are in a separate outer field of NGC 6752. The nine
  objects indicated by blue numbers are initially selected CV
  candidate counterparts that fall in this magnitude range. The CX2,
  CX4, CX5, CX7, CX21, and CX35 counterparts are clearly consistent
  with cluster membership, while the CX3, CX10, and CX15 counterparts
  have a discordant proper motion that is consistent with the mean
  galaxy proper motion. We thus interpret these sources as being
  likely AGNs, as discussed in the text. The CX29 counterpart (not
  shown) also has a discordant proper motion; however this may well be
  compromised by the presence of a much brighter neighbor (see
  Fig.~\ref{f:finding_charts}).
  \label{f:PM}}
\end{figure*}

\begin{figure*}
\centering
\includegraphics*[clip,viewport=18 144 592 718,width=5.5in]{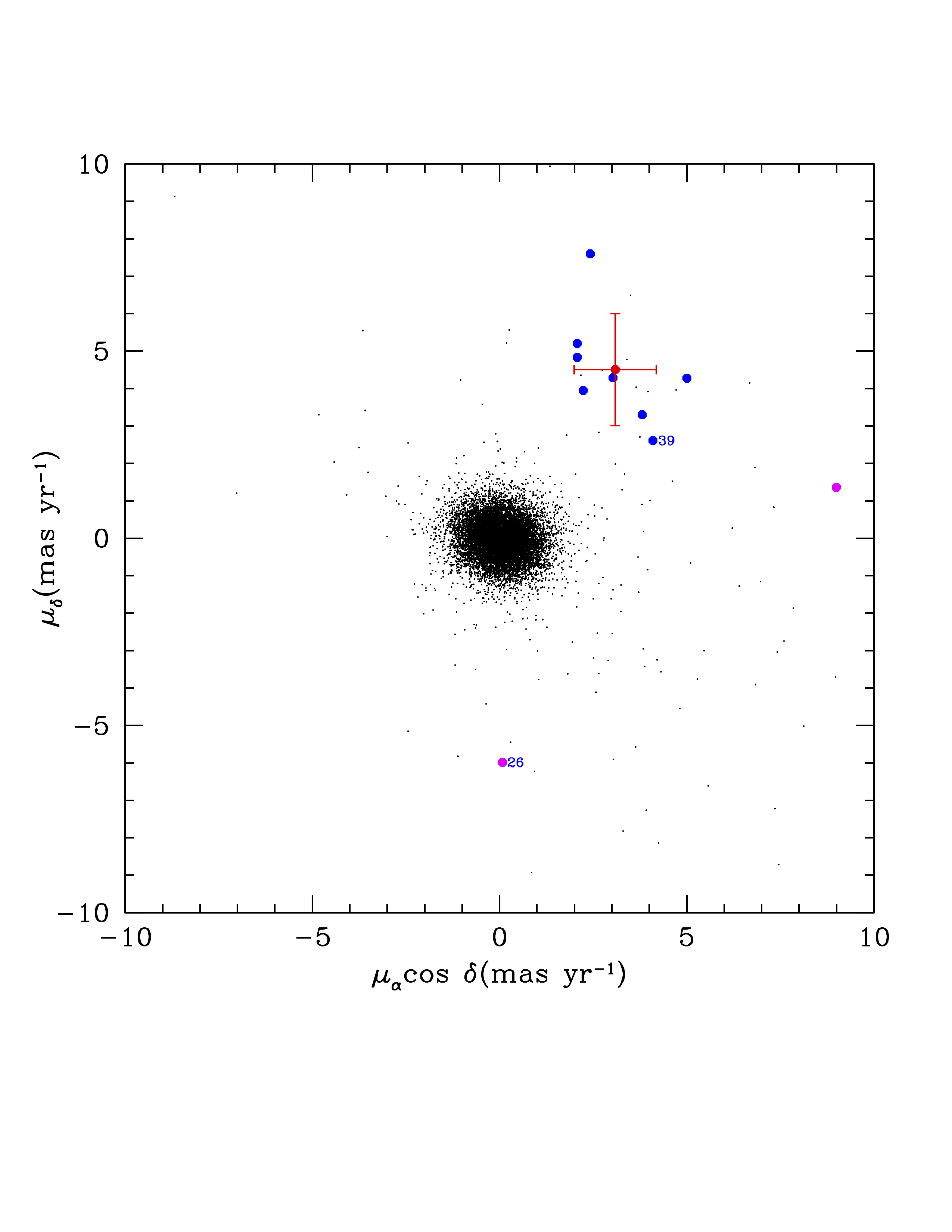}
\figcaption{Proper-motion component distribution for stars in the
  magnitude range $21 \le \R < 23$ with the proper motions of 10
  galaxies overplotted as blue and magenta dots.  The zero-point
  corresponds to the systemic cluster motion. The red point and error
  bars indicate the mean and standard error of the mean of the eight
  brightest galaxies (blue dots), six of which are in the mosaic field
  and two of which are in a separate outer field of NGC 6752. The two
  faintest galaxies (magenta dots) have discordant proper motions
  relative to the clump defined by the eight brightest galaxies and
  thus were not included in the average. The two numbered dots are the
  candidate counterparts to sources CX26 and CX39 (see
  Fig.~\ref{f:finding_charts}). 
  \label{f:galaxy_PM}}
\end{figure*}

\subsection{Astrometry\label{astrometry}}

The drizzle-combined \hst\ ACS/WFC mosaic for each filter was
rectified to the ICRS using approximately 600 astrometric standards
from the USNO UCAC3 catalog.  The RMS residual of the plate solution
was 0\farcs09 in each coordinate.  We determined a boresight
correction for the \chandra\ source coordinates from
\citet{Forestell14} by computing the mean offsets between the
\hst\ and \chandra\ coordinates for three of the brightest CVs from
\citet{Pooley02}, viz.\ sources CX2, CX3, and CX4.  The total shift in
the coordinates between the \chandra\ and \hst\ systems was
approximately 0\farcs1.

We searched for optical counterparts to the \chandra\ sources by
overlaying the \chandra\ error circles on the drizzle-combined
\hst\ image mosaics, with the boresight correction applied to the
\chandra\ source positions from \citet{Forestell14}.  Since the
uncertainty in the optical positions $(\lesssim 0\farcs1$) was small
compared with the uncertainty in the X-ray positions ($\lesssim
0\farcs3$), we neglected the contribution of the former to the total
positional uncertainty.

In order to test candidate counterparts for cluster membership, we
computed proper motion components for all of the objects in the region
covered by the image mosaic.  The present dataset was used for the
reference epoch.  We used the GO-10775 (PI: Sarajedini) dataset for
the second epoch.  It was obtained with the ACS/WFC in 2006 May and
covers very nearly all of the half-mass radius region.  With a mean
epoch for our dataset of 2011 Aug, the two epochs provide a 5.3-year
time baseline for proper motion determinations.

The distribution of proper motion components depends on the magnitudes
of the stars for which the proper motions are measured, since
measurement uncertainties become dominant for the fainter stars.
Thus, we compute the amplitude of the proper motion for each of the
candidate counterparts and compare it to the RMS proper motion
amplitude for stars of similar magnitude.  We judge membership to be
unlikely for stars with proper motion amplitudes that exceed the RMS
amplitude by more than a factor of three.  Figure~\ref{f:PM} shows the
proper motion component distribution for stars in the wide magnitude
range $18 \le \R < 23$, with the nine initially selected CV candidates
in this range with measured proper motions indicated.  Six of the
candidates have proper-motion amplitudes that are clearly consistent
with membership, while three---the counterparts to CX3, CX10, and
CX15---are clearly discordant. It is striking that these three objects
have nearly the same proper motion, suggesting that they might, in
fact, be background active galactic nuclei (AGNs).

To investigate this possibility, we determined the absolute proper
motion of NGC 6752 by locating a sample of galaxies in our image
mosaic and also in an outer field that was imaged by the ACS/WFC in
2004 and 2012. We used the SExtractor software \citep{Bertin96} to
detect objects in both fields and filtered the list of detected
objects on a combination of ``stellarity'' index, ellipticity, PSF
FWHM, and the difference between the maximum surface brightness and
the Kron magnitude\footnote{This difference measures the central
  concentration of the object image and provides another means of
  star/galaxy discrimination \citep{Annunziatella13}.} to generate a
list of candidate galaxies. This list includes both galaxies and
bright-star PSF artifacts. In order to filter out the later, we
visually inspected each of the several hundred galaxy candidates and
selected the obvious galaxies. This resulted in a list of 10 galaxies
with measured proper motions, eight in the central region and two in
the outer region. The result of this analysis is shown in
Figure~\ref{f:galaxy_PM}, where the galaxy proper motions are
overplotted on the distribution of proper motions for all objects in
the magnitude range $21 \le \R \le 23$.  With the exception of two
outliers, the galaxies fall in a compact clump, which defines the
absolute proper-motion zero-point. The two outliers are, in fact, the
faintest two galaxies in the sample and thus likely have the least
reliably determined proper motions. The mean galaxy proper motion is
plotted in both Figs.~\ref{f:PM} and \ref{f:galaxy_PM}. It is clear
from these figures that the proper motions of the proposed
counterparts to CX3, CX10, and CX15 agree with the galaxy mean. Given
this agreement in proper motion, it appears highly likely that CX3,
CX10, and CX15 are actually background active galactic nuclei (AGNs)
rather than CVs in the cluster. Further evidence for this
interpretation is given in \S\ref{source_types}.

\section{Results}

\subsection{\chandra\ Source Identification \label{source_ID}}

We used the optical CMDs to detect likely \chandra\ source
counterparts and investigate their properties.  For each of the 39
\chandra\ sources within the half-mass radius, we checked the CMD
locations of all objects that fell within a distance of the maximum of
2.5 times the error circle radius and 0\farcs3.  The rationale for
choosing this search region size is that the formal {\tt pwdetect}
error circle radii are quite small for the brightest sources ($\sim
0\farcs02$) and in some of these cases potential candidates were
located near, but not within, the actual error circle.  We note that
\citet{Hong05} have observed that wavelet detection algorithms
systematically underestimate positional uncertainty.  Their
prescription for determining 95\% positional uncertainty produces an
asymptotic lower-limiting value of about 0\farcs3 for an on-axis
source.

\newcommand{\nd}{\nodata}
\newcommand{\rr}{\nointerlineskip\raggedright\hangindent 1.5ex \hangafter 1}
\newcommand{\notebox}[1]{\parbox[t]{1.25in}{\rr #1\vskip 2pt}}
\newcommand{\tnmi}{\tablenotemark{i}}
\tabletypesize{\scriptsize}
\startlongtable
\begin{deluxetable*}{llcccclccccl}
\tablecolumns{13}
\tablewidth{8in}
\tablecaption{\textbf{Optical Counterpart Summary}\label{t:counterparts}}
\tablehead{
\colhead{Source\tablenotemark{a}} &
\colhead{RA, Dec (J2000)\tablenotemark{b}} &
\colhead{$r_{\rm err}~('')$\tablenotemark{c}} &
\colhead{$r~(')$\tablenotemark{d}} &
\colhead{$N_{\rm detect}$\tablenotemark{e}} &
\colhead{Offset\tablenotemark{f}} &
\colhead{Type\tablenotemark{g}} &
\colhead{PM\tablenotemark{h}} &
\colhead{\R} &
\colhead{\B} & 
\colhead{$\ha$} &
\colhead{Notes}
}
\startdata
%
%
CX1    &  19:10:51.138  $-$59:59:11.92  &  0.01  &  0.17  &  2       &  1.8  &  CV   & \nd &  19.38  &  19.46  &  19.01  &  \\                                        
CX2    &  19:10:56.005  $-$59:59:37.33  &  0.02  &  0.73  &  1       &  1.0  &  CV   & c   &  19.22  &  20.36  &  18.60  &  \\
CX3    &  19:10:40.375  $-$59:58:41.47  &  0.02  &  1.52  &  1       &  1.3  &  AGN  & f   &  20.97  &  21.76  &  20.72  &  \\
CX4    &  19:10:51.586  $-$59:59:01.73  &  0.02  &  0.08  &  2       &  0.8  &  CV   & c   &  20.10  &  21.09  &  19.35  &  \\
CX5    &  19:10:51.414  $-$59:59:05.18  &  0.02  &  0.09  &  1       &  1.6  &  CV?  & c   &  18.57  &  19.65  &  18.33  &  \\
CX6    &  19:10:51.505  $-$59:59:27.10  &  0.03  &  0.38  &  1       &  0.8  &  CV?  & \nd &  23.87  &  24.07  &  23.78  &  \notebox{very blue, small \ha\ excess} \\
CX7    &  19:10:51.511  $-$59:58:56.85  &  0.02  &  0.15  &  2       &  1.9  &  CV   & c   &  20.94  &  22.01  &  19.91  &  \\
CX8    &  19:11:02.969  $-$59:59:41.92  &  0.05  &  1.49  &  3       &  5.3  &  AB?  & \nd &  18.34  &  20.91  &  17.54  &  \notebox{two red, \ha-excess objects} \\
CX8    &                                &        &        &  3\tnmi  &  6.4  &  AB?  & \nd &  18.48  &  21.06  &  17.67  &  \\
CX9    &  19:10:51.766  $-$59:58:59.25  &  0.04  &  0.10  &  2       &  3.0  &  CV?  & \nd &  22.42  &  24.40  &  22.01  &  \notebox{somewhat blue, \ha\ excess in color-color diagram} \\ 
CX10   &  19:10:54.754  $-$59:59:13.86  &  0.06  &  0.37  &  1       &  1.2  &  AGN  & f   &  19.70  &  20.16  &  19.47  &  \\
CX11   &  19:10:52.408  $-$59:59:05.64  &  0.08  &  0.04  &  2       &  \nd  &  \nd  & \nd &  \nd    &  \nd    &  \nd    &  \notebox{counterpart suggested by \citet{Pooley02} is not detected; only MS stars present in search area; MSP D} \\
CX13   &  19:10:40.610  $-$60:00:05.91  &  0.14  &  1.76  &  1       &  1.0  &  CV   & c   &  24.32  &  24.60  &  23.38  &  \\
CX14   &  19:10:52.063  $-$59:59:09.09  &  0.08  &  0.08  &  2       &  \nd  &  \nd  & \nd &  \nd    &  \nd    &  \nd    &  \notebox{only MS stars present in search area} \\
CX15   &  19:10:55.847  $-$59:57:45.78  &  0.08  &  1.39  &  1       &  1.0  &  AGN  & f   &  22.66  &  23.05  &  22.41  &  \\
CX16   &  19:10:42.531  $-$59:58:43.03  &  0.13  &  1.25  &  1       &  1.1  &  AB   & c   &  18.33  &  19.39  &  18.09  &  \\
CX17   &  19:11:05.258  $-$59:59:04.42  &  0.16  &  1.65  &  1       &  \nd  &  GLX  & \nd &  \nd    &  \nd    &  \nd    &  \notebox{asymmetric, extended object; photometry not possible} \\
CX18   &  19:10:52.056  $-$59:59:03.71  &  0.14  &  0.02  &  2       &  \nd  &  \nd  & \nd &  \nd    &  \nd    &  \nd    &  \notebox{only MS stars present in search area} \\
CX19   &  19:10:55.600  $-$59:59:17.33  &  0.16  &  0.49  &  3       &  1.8  &  AB?  & c   &  17.58  &  18.35  &  17.37  &  \notebox{normal \br, small \ha\ excess} \\
CX20   &  19:10:52.848  $-$59:59:02.54  &  0.10  &  0.10  &  1       &  2.7  &  AB   & c   &  20.20  &  21.99  &  19.87  &  \\
CX21   &  19:10:49.516  $-$59:59:43.16  &  0.11  &  0.72  &  2       &  0.7  &  CV?  & c   &  18.53  &  19.06  &  18.40  &  \notebox{blue, small \ha\ excess in color-color diagram} \\
CX22   &  19:11:02.950  $-$59:57:58.91  &  0.13  &  1.74  &  0       &  \nd  &  \nd  & \nd &  \nd    &  \nd    &  \nd    &  \notebox{empty search region} \\
CX23   &  19:10:52.546  $-$59:59:04.38  &  0.11  &  0.06  &  4       &  0.4  &  CV?  & c   &  19.58  &  20.68  &  19.11  &  \notebox{uncertain photometry} \\ 
CX24   &  19:10:52.670  $-$59:59:03.21  &  0.12  &  0.07  &  4\tnmi  &  3.4  &  CV?  & \nd &  22.66  &  22.97  &  22.61  &  \notebox{weakly detected in \R\ and \ha} \\ 
CX25   &  19:10:51.957  $-$59:58:40.55  &  0.13  &  0.40  &  5\tnmi  &  3.9  &  CV?  & \nd &  25.51  &  25.73  &  24.81  &  \notebox{weakly detected in \R\ and \ha} \\ 
CX26   &  19:10:39.162  $-$59:59:45.15  &  0.23  &  1.75  &  2       &  1.9  &  GLX  & f   &  23.00  &  25.16  &  22.55  &  \notebox{extended elliptical image} \\
CX27   &  19:10:52.059  $-$59:59:00.84  &  0.12  &  0.06  &  1\tnmi  &  3.7  &  AB?  & c   &  20.13  &  21.65  &  19.83  &  \notebox{normal \br, small \ha\ excess; MSP B} \\
CX28   &  19:10:42.509  $-$59:59:44.46  &  0.22  &  1.37  &  0       &  \nd  &  \nd  & \nd &  \nd    &  \nd    &  \nd    &  \notebox{empty search region} \\
CX29   &  19:10:52.293  $-$59:59:01.79  &  0.16  &  0.05  &  4       &  0.4  &  CV?  & f   &  20.83  &  22.49  &  20.50  &  \notebox{slightly blue, small \ha\ excess} \\
CX30   &  19:10:40.678  $-$59:58:39.61  &  0.19  &  1.49  &  0       &  \nd  &  \nd  & \nd &  \nd    &  \nd    &  \nd    &  \notebox{faint object in wings of much brighter star; not detected by KS2} \\
CX31   &  19:10:50.514  $-$59:57:37.11  &  0.17  &  1.47  &  1       &  \nd  &  \nd  & \nd &  \nd    &  \nd    &  \nd    &  \notebox{only red giant present in search area} \\
CX32   &  19:10:54.137  $-$59:59:11.04  &  0.17  &  0.28  &  1       &  1.7  &  CV?  & \nd &  20.54  &  22.12  &  20.23  &  \notebox{slightly blue, slight \ha\ excess} \\
CX33   &  19:11:03.287  $-$59:58:01.31  &  0.19  &  1.75  &  0       &  \nd  &  \nd  & \nd &  \nd    &  \nd    &  \nd    &  \notebox{empty search region} \\
CX34   &  19:10:45.694  $-$59:58:20.09  &  0.20  &  1.09  &  1       &  1.6  &  AB   & c   &  20.23  &  22.04  &  19.88  &  \\
CX35   &  19:10:52.165  $-$59:59:16.73  &  0.19  &  0.20  &  2       &  0.4  &  CV?  & \nd &  24.11  &  24.89  &  23.35  &  \notebox{uncertain photometry} \\
CX36   &  19:10:49.585  $-$59:58:26.41  &  0.20  &  0.71  &  1       &  2.2  &  CV   & \nd &  24.85  &  24.93  &  23.81  &  \\
CX37   &  19:10:50.509  $-$59:59:08.77  &  0.21  &  0.21  &  4       &  1.1  &  AB?  & c   &  20.42  &  22.35  &  20.12  &  \notebox{somewhat red, no \ha\ excess in color-color diagram} \\
CX38   &  19:11:02.151  $-$59:58:11.81  &  0.24  &  1.53  &  1       &  1.4  &  AB?  & c   &  19.15  &  20.34  &  18.93  &  \notebox{slightly red, slight \ha\ excess} \\
CX39   &  19:10:46.352  $-$59:57:49.92  &  0.22  &  1.44  &  1       &  1.8  &  GLX  & f   &  22.26  &  24.86  &  21.89  &  \notebox{extended elliptical image} \\
CX40   &  19:10:50.357  $-$59:59:13.89  &  0.27  &  0.27  &  7       &  2.2  &  AB   & c   &  19.72  &  21.18  &  19.45  &  \\
\enddata			  
\tablenotetext{a}{From \citet{Forestell14}.}
\tablenotetext{b}{\chandra\ source positions from \citet{Forestell14}
  have been boresight-corrected to align with the drizzled image
  coordinate system, \\ which is tied to the ICRS.}
\tablenotetext{c}{Wavdetect error circle radius in arcsec. The search
  area radius is $\max(2.5\, r_{\rm err},0.3'')$.}
\tablenotetext{d}{Projected distance from cluster center in arcmin.}
\tablenotetext{e}{Number of objects detected within $\max(2.5\, r_{\rm err},0.3'')$.}
\tablenotetext{f}{Offset of counterpart from X-ray position in units of $r_{\rm err}$.}
\tablenotetext{g}{%
CV = cataclysmic variable; 
CV? = less certain CV identification for reason noted in table;
AB = active binary candidate; \\ 
AB? = less certain AB identification for reason noted in table; 
GLX = galaxy;
AGN = active galactic nucleus.
}
\tablenotetext{h}{Proper-motion membership: 
  c = consistent with  cluster; 
  f = consistent with field; 
  \nd\ = no proper-motion measurement.}
\tablenotetext{i}{Preferred counterpart lies somewhat outside of formal search region.}
\end{deluxetable*}

\begin{figure*}
\epsscale{0.92}
\plotone{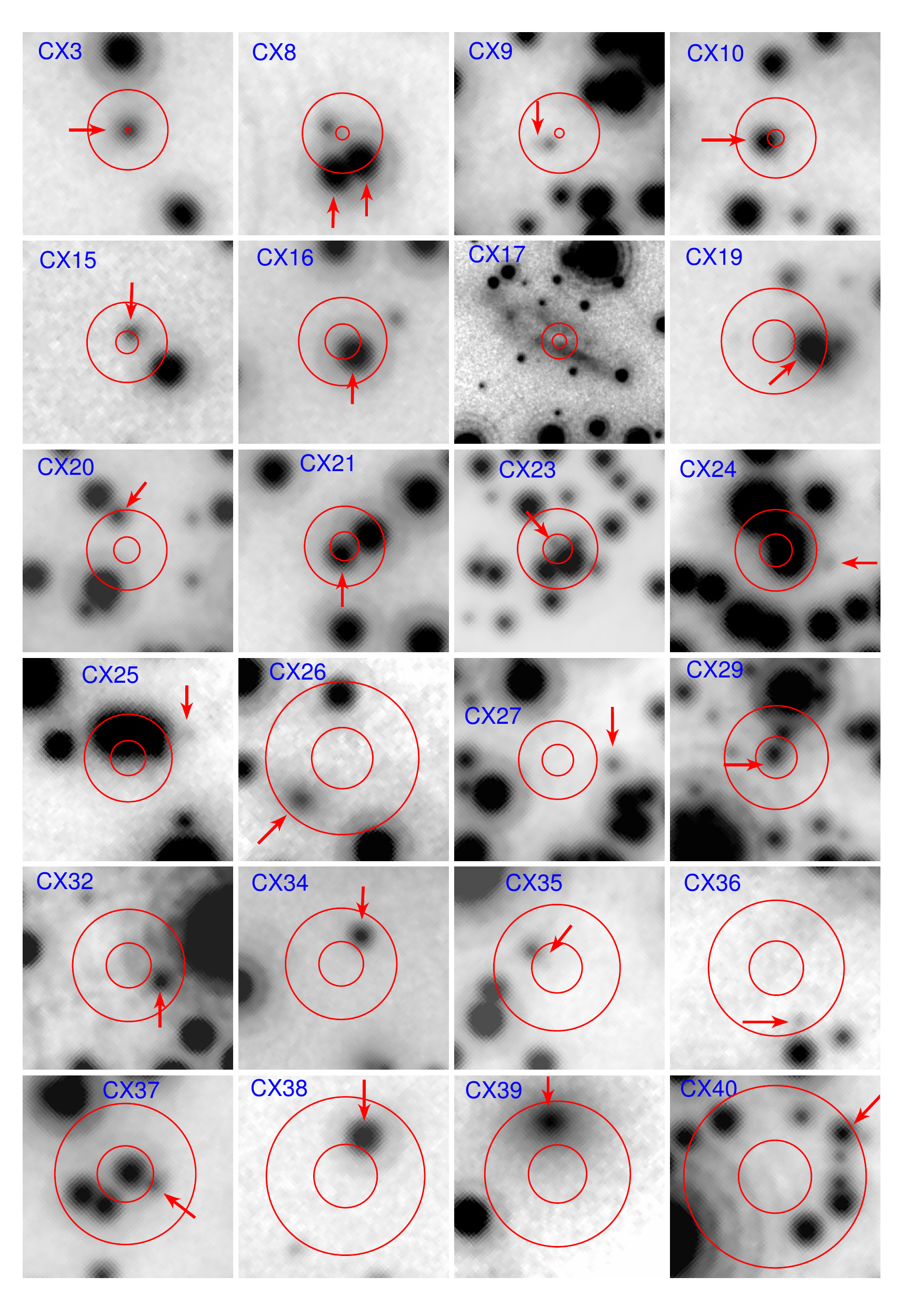}
\figcaption{Finding charts for revised and new source
  identifications. All charts are for the \ha\ band, with the
  exception of those for sources CX24 and CX25, which are for the
  \B\ band, in which the counterpart was more strongly detected. North
  is up and east is to the left. With the exception of the chart for
  CX17, the regions shown are $1\farcs5 \times 1\farcs5$. For
  CX17, the region is $4\farcs5 \times 4\farcs5$.  The inner
  circles represent the formal error circle and the outer circles
  represent the search regions, which have a radius of the maximum of
  $2.5\, r_{\rm err}$ and 0\farcs3. The arrows point to the
  candidate counterparts. As discussed in \S\S\ref{astrometry} and
  \ref{source_types}, the counterparts to CX3, CX10, and CX15, which
  were previously classified as CVs, now appear to be AGN, based on
  their proper motions and lack of an \ha\ excess.
  \label{f:finding_charts}}
\end{figure*}

Objects that fell on the main sequence, subgiant branch, or giant
branch were considered to be unlikely counterparts, given the
relatively low X-ray to optical flux ratio, $f_X/f_{\rm opt}$, of
such stars, in contrast to the ranges for chromospherically active
binaries and cataclysmic variables.  Table~\ref{t:counterparts}
summarizes the result of this counterpart
search. Figure~\ref{f:finding_charts} shows finding charts for
identifications that have changed or are new since the previous
studies by \citet{Pooley02} and \citet{Thomson12}.

\subsection{Source Types \label{source_types}}

Based on the location of the proposed counterparts in the CMDs, we
primarily assigned types of candidate cataclysmic variable (CV) and
candidate chromospherically active binary (AB)\@.  Candidate CVs were
generally identified by being significantly to the blue of the MS
and/or having large \ha\ excesses (either relative to the MS or to the
WD sequence).  ABs were defined as lying within $\sim0.75$~mag above
the MS (and thus within $\sim0.2$~mag to the red of the MS) and having
small \ha\ excesses ($\la 0.1$~mag), based on our previous analysis of
ABs in NGC~6397 \citep{Cohn10}.

Although our proposed counterpart to CX5 lies slightly to the red side
of the MS, its high $L_X = 1.1\times10^{32}~\ergs$ and moderately high
$f_X/f_{\rm opt} = 1.4$ tend to support its identification as a CV.\@
One possibility is that the object detected at the position of CX5 is
a MS star that is covering up the much fainter true counterpart.
However, examination of the stellar image does not indicate any
obvious asymmetry. Given the weak \ha~excess of CX5, we include it
among the less certain CV identifications.

In cases where only MS stars were present in the search region
(viz.\ CX11, CX14, CX18, and CX35) and in one case where a red giant
was present in the search area (CX31), we noted this in
Table~\ref{t:counterparts} and did not assign a counterpart.  We note
that an AB with a low mass ratio and weak lines could look like a MS
star in both CMDs.  In some of these cases, it may be that a bright
star near the \chandra\ source location is covering up a much fainter
star.

Three objects, viz.\ sources CX17, CX26, and CX39, were classified as
background galaxies, based on the extended appearance of their images.
The \R-band images of the source CX26 and CX39 counterparts resemble
elliptical galaxies in appearance, with a smooth elongation.  The
image of the source CX17 counterpart has a more complex structure,
possibly suggestive of interacting galaxies.  The two apparent
elliptical galaxies have moderate X-ray to optical flux ratios
of $\sim0.3 - 1$, consistent with normal galaxies (see
\S\ref{flux_ratio}).  It was not possible to perform PSF-fitting
photometry on the source CX17 counterpart, given its rather amorphous
structure, without a central nucleus, and thus we did not compute an
X-ray to optical flux ratio for it.  As noted by \citet{Pooley02},
this source coincides with a radio source, which further supports its
identification as a galaxy.

Three objects that have been previously classified as CVs,
viz.\ sources CX3, CX10, and CX15, were ultimately classified as
background AGNs, based on their proper motions, which are discordant
from the cluster distribution but in agreement with the galaxy mean
(which included the proper motion of the counterpart to
CX39).\footnote{The counterpart to CX26 has a discordant proper motion
  from the eight galaxies used to calculate the galaxy mean. We
  attribute this to a poorly determined proper motion for this object.}
The deep imaging in this study indicates that these three source
counterparts are pointlike, without any hint of extension. This is
consistent with an AGN interpretation. As can be seen in
Fig.~\ref{f:fx_fopt}, these three sources have moderate to high X-ray
to optical flux ratios of $\sim0.3 - 20$, consistent with
AGNs. Examination of the (\hr, \R) CMD indicates that the counterparts
to these three sources do not show an \hr\ excess although all three
counterparts are blue in the (\br, \R) CMD\@. This is further evidence
for the AGN interpretation, as AGN would not be expected to show a
zero-redshift \ha\ excess.

Figure~\ref{f:CMD_CV} shows the location of the likely or less certain
CV candidates in the CMDs.  As we noted in \citet{Cohn10} for the CV
population in NGC~6397, there is a brighter group of CV candidates
that generally lie blueward of the main sequence in \br\ and generally
show \ha\ excesses, and a fainter group of CV candidates that are
distributed around the WD cooling sequence in the (\br, \R) CMD and
generally show \ha\ excesses relative to the WD sequence.  All but two
of the 11 brightest CV candidates lie 0.1 -- 1.3 mag to the blue of
the MS in the (\br, \R) CMD; the candidate counterpart to source CX2
lies on the MS and the counterpart to CX5 lies slightly to the red of
the MS\@.  As we noted in \citet{Cohn10}, the optical emission of the
bright CV systems appears to be dominated by the secondary in the
\R\ band, with a larger contribution from the disk in the
\B\ band. The fairly high \R-band fluxes indicate that the secondaries
are relatively massive, $\sim 0.5-0.7~\msun$.  Five of the nine
brightest CV candidates, viz.\ sources CX1, CX2, CX4, CX7, and CX23,
show \ha\ excesses of 0.2 -- 0.8 mag relative to the MS\@.  Of the
remaining four bright CV candidates, sources CX5, CX29, and CX32 show a
small \ha\ excess, and source CX21 is slightly to the \ha\ deficit
side of the MS\@. Though CX21 does not show a \ha\ excess, its
proper motion verifies that it is a cluster member, and therefore
cannot be an AGN. Further investigation of the \ha-status of the
proposed CV counterparts using the color-color diagram is described
below.

About 1.5 mag below the faintest of the bright CV candidates are two
possible transitional objects, the counterparts to sources CX9 and
CX24\@. The candidate counterpart to CX9 has a small \br\ excess and
lies on the MS in the (\hr, \R) CMD\@. The candidate counterpart to
CX24 lies near the WD regions in both the (\br, \R) and (\hr, \R)
CMDs\@.

About 1.5 mag below the proposed CX9 and CX24 counterparts lie
five faint CV candidates. These fall on or near the WD sequence in
the (\br, \R) CMD\@.  Four of the five, viz.\ the counterparts to
sources CX13, CX25, CX35, and CX36, show \ha\ excesses relative to the
normal WD sequence in the (\hr, \R) CMD\@.  The fourth faint CV
candidate, source CX6, is found among the main distribution of white
dwarfs in its \hr\ color index.  As we noted in \citet{Cohn10}, the
optical fluxes for the faint CV candidates in NGC~6397 are dominated
by the contribution of the WD\@.  The \ha\ excess that four of the
five faint CV candidates show relative to the WD distribution suggests
that the faint CVs have a strong \ha-emission core (due to an
accretion disk) within the broad absorption lines of the WD continuum.

The five faint CV candidates in NGC~6752 are clustered near $\R \sim
24.5$, which corresponds to $M_R \sim 11.4$, which is similar to the
magnitude of the clump of faint CVs observed in NGC~6397.  As we
noted, this characteristic magnitude is also similar to that of the
sharp peak in the field CV distribution observed in the Sloan Digital
Sky Survey by \citet{Gaensicke09}.  The peak in the field distribution
occurs near the CV period minimum ($P\approx \mbox{80 -- 86}\,{\rm
  min}$).  Thus, it appears likely that the faint CVs that we are
observing in NGC~6752 also correspond to the period-minimum
population. We can also compare the X-ray luminosity to detailed
simulations of CV evolution in globular clusters. We use the
simulations of \citet{Ivanova06}, which predict the optical and X-ray
flux of CVs in globular clusters throughout their evolution. In the
range $M_V = 11-12$, we find that 80\% of their simulated CVs are
predicted to have periods below the period gap, but not yet increasing
in period, with small numbers of CVs above the period gap,
``period-bouncers'' with degenerate hydrogen-rich companions, and AM
CVn stars. The predicted $L_X$ range for these populations is
$L_X(0.5-8\,{\rm keV}) = 5\times10^{29}-3\times10^{31}\,\ergs$, and
the median $L_X$ is $\sim 4\times10^{30}\,\ergs$; these well match the
$L_X$ range of these five candidate faint CVs,
$7\times10^{29}-5\times10^{31}\,\ergs$, and their median of
$2\times10^{30}\,\ergs$ (from \citealt{Forestell14}).

\begin{figure*}
\centering
\includegraphics*[clip,viewport=18 144 592 718,width=5.5in]{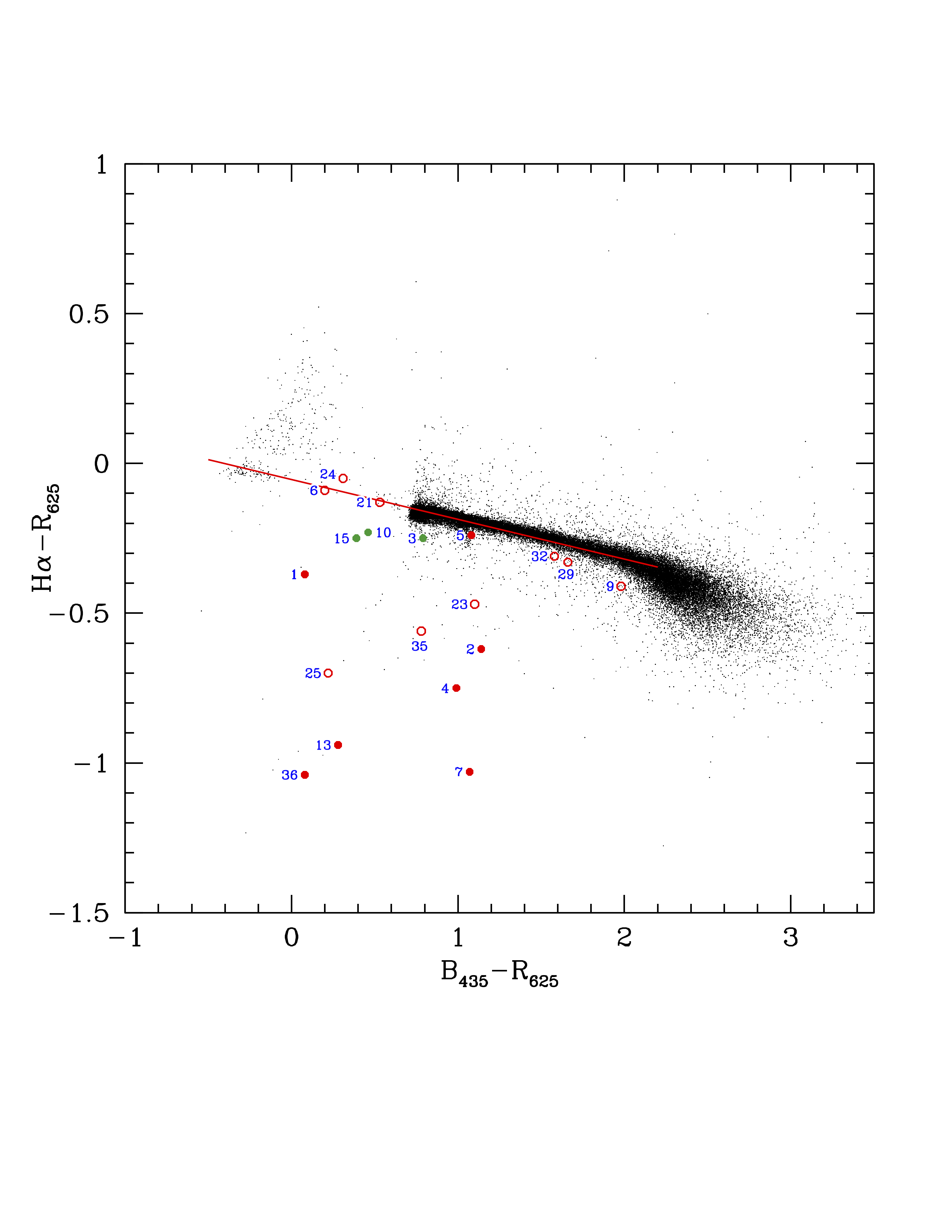}
\figcaption{Proper-motion cleaned color-color diagram for stars within
  the half-mass radius of NGC~6752 and CV identifications.  The
  candidates are the same as in Fig.~\ref{f:CMD_CV}.  Open symbols
  indicate less certain CV identifications.  The red line is a linear
  regression of \hr\ on \br\ over the range $-0.5 \le \br \le
  2.2$. All stars brighter than $\R = 15.5$ have been excluded, since
  saturation effects set in at the bright end.  The blue end of the
  color-color relation is populated by stars on the extreme blue
  horizontal branch. Note that all of the candidates except the
  counterpart to CX24 lie below (i.e.\ to the \ha-excess side of) the
  color-color relation.  
\label{f:CCD_CV}}
\end{figure*}

\begin{figure*}
\centering
\includegraphics*[clip,viewport=18 144 592 718,width=5.5in]{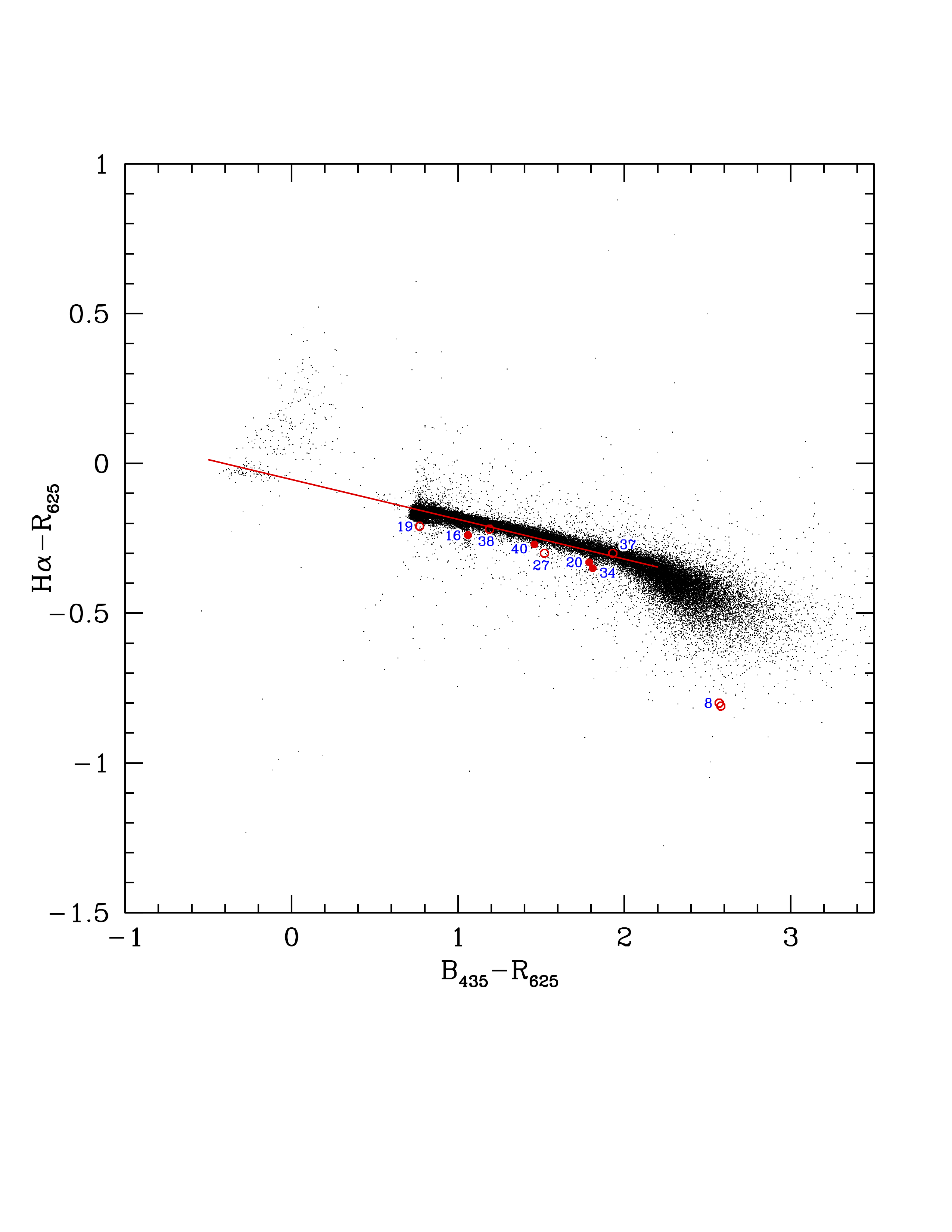}
\figcaption{Proper-motion cleaned color-color diagram for stars within
  the half-mass radius of NGC~6752 and AB identifications.  The
  candidates are the same as in Fig.~\ref{f:CMD_AB}.  Note that all of
  the candidates except the counterpart to CX37 lie below (i.e.\ to
  the \ha-excess side of) the color-color relation.  
\label{f:CCD_AB}}
\end{figure*}

Further insight into the \ha-status of the CV candidates NGC~6752 is
provided by the color-color diagram, Fig.~\ref{f:CCD_CV}. It can be
seen that there is a nearly linear relation between the two colors, as
indicated by the red line that represents a linear regression of \hr\ on
\br\ over the range $-0.5 \le \br \le 2.2$. All of the proposed
counterparts, with the exception of those to sources CX6, CX21, and
CX24, fall significantly below the relation. Thus, in some cases where
the candidate does not lie to the \ha-excess side of the MS in the
(\hr, \R) CMD, it still shows a significant \ha-excess relative to
other objects of its \br\ color. The proposed counterparts to CX6 and
CX21 fall very slightly to the \ha-excess side of the color-color
relation, while the proposed counterpart to CX24 falls a small
distance to the \ha-deficit side of the relation, though all three
counterparts are consistent with the MS color-color relation, within
the \hr\ scatter at their \R\ magnitudes ($\sim0.1$\,mag at CX24's
$\R=22.66$).

We can think of several scenarios to explain these systems without
clear \ha\ excesses or deficits. There might be only small
\ha\ emission, which is not strong enough to dominate over the
\ha\ absorption. For example, CX21 might be interpreted as a bright,
nova-like CV, with an accretion flow that is mostly optically thick,
leading to weak \ha\ emission. Perhaps there is no accretion disk,
which would suggest a radio millisecond pulsar nature. CX24 indeed has
a small X-ray luminosity and soft X-ray color, consistent with the
X-rays from most radio millisecond pulsars
(e.g.\ \citealt{Bogdanov11}). However, CX6 is brighter and has a
harder X-ray spectrum. This is consistent with radio millisecond
pulsars showing strong shocks, the redbacks and black widows
(e.g.\ \citealt{Gentile14,Bogdanov10}), but the optical data show a He
WD companion, apparently ruling out a millisecond pulsar nature for
CX6. CX21's optical counterpart, however, is consistent with a redback
millisecond pulsar, as is its low X-ray luminosity and moderately soft
hardness. Finally, any or all of these three objects could be AM CVn
stars, where a He WD (that has lost its outer H layer) donates mass to
(typically) a CO WD\@. In such a system, optical flux could come from
the accreting WD and/or the accretion disk, neither of which would
show hydrogen lines in emission or absorption. Such systems have been
predicted to be present in globular clusters (e.g.\ by
\citealt{Ivanova06}) and evidence for the detection of an AM CVn
binary in NGC 1851 has been presented by \citet{Zurek16}.

Figure~\ref{f:CMD_AB} shows the location of the objects identified as
AB stars in the CMDs\@.  As in NGC~6397, these objects sometimes lie
close to the edge of the MS in either the right or left panel, but
deviate from the MS by a larger amount in the other panel.  Of the
nine AB counterparts listed in Table~\ref{t:counterparts}, eight are
newly identified in our study.  This increase in the detection rate
appears to result both from the deeper \chandra\ imaging dataset that
we used, compared with that of \citet{Pooley02}, and the increased
photometric precision possible with the ACS/WFC vs.\ the WFPC2, which
allows us to identify objects that deviate by small amounts from the
MS\@.  As seen in NGC~6397 and 47~Tuc, the ABs are likely to be a
mixture of BY Draconis stars, W Ursa Majoris stars, and other contact
binaries \citep[see][]{Albrow01,Taylor01,Cohn10}.

Figure~\ref{f:CCD_AB} shows the location of the objects identified as
AB stars in the color-color diagram. As can be seen in the figure, all
of the AB candidates except for the counterpart to CX37 fall below the
line that represents the linear regression to the color-color
relation. As in NGC~6397, the \ha\ excesses of the AB candidates are
typically much smaller than those of the CV candidates. This is
consistent with the findings of \citet{Kafka06} and \citet{Beccari14}
that chromospherically active binaries do not show \ha\ equivalent
widths in excess of about 10\,\AA\@. An apparent exception is the
object pair that is the proposed counterpart to CX8, which falls well
below the linear regression, as discussed and explained below.

The faintest ABs that are detected in the X-ray in NGC~6752 have $\R
\approx 20.5$, corresponding to $M_R \approx 7.4$.  At this magnitude,
we reach the X-ray detection limit of $L_X \approx 5 \times
10^{29}~\ergs$.  In contrast, the AB sequence that we detected in
NGC~6397 reaches a deeper limit of $M_R \approx 9.7$, owing to the
deeper X-ray detection limit of $L_X \approx 9 \times 10^{28}~\ergs$.
Given deeper X-ray imaging, it should be possible to trace the AB
sequence farther down the MS of NGC~6752. New, deeper
\chandra\ observations are now scheduled for 2017.

Our proposed counterpart to source CX8 stands out for its very red
color and large apparent \ha\ excess.  There are actually two
photometrically similar stars located 0\farcs2 apart, within about
0\farcs3 of the X-ray source position (see
Fig.~\ref{f:finding_charts}; the photometry for these stars is plotted
in Fig.~\ref{f:CMD_AB}). We found similarly discordant stars in
NGC~6397 and interpreted them as likely foreground ABs superposed on
the cluster, with distance moduli that put them well above the
fiducial sequences in the CMDs \citep[][see their \S4.2 and
  Fig.~4]{Cohn10}.  The vertical offset for the possible CX8
counterparts, roughly estimated at $3.6-4.6$ magnitudes (taking into
account the opposite effects of metallicity and reddening on NGC~6752
stars), suggests a distance $5-8$ times closer, or $500-800$\,pc.  At
this distance, the two stars are $100-160$\,AU in projection from each
other, making them likely to be bound, but the separation is much too
large to produce chromospheric activity. One possible explanation for
the activity is that the two stars are both unresolved binaries. A
plausible explanation (which we explore further below) is that the two
stars are each an unresolved binary. (We note that explaining the
activity as due to youth is rather unlikely, since only a tiny
fraction of field M stars are likely to be young enough to have
substantial chromospheric activity, while many M stars are likely to
reside in tight binaries.)

We note that even without actual \ha\ emission, M star spectra have a
peak at the location of the \ha\ line, owing to the presence of TiO
bands to either side of this wavelength. We have used the {\tt
  synphot} utility in {\small IRAF/STSDAS} to estimate the apparent
\hr\ excess for a normal M5V star as a function of metallicity. We
used the \citet{Castelli04} spectra for metallicity values of $-1.5$
(appropriate to NGC~6752), 0.0 (solar), and 0.5 (super solar), with
$T_{\rm eff} = 3500\,{\rm K}$ and $\log(g) = 5.0$. The predicted
values of \hr\ for these three metallicities are $-0.41$, $-0.58$, and
$-0.59$, respectively. We note that our proposed counterparts to CX8
have a \hr\ value of $-0.80$, suggesting that there is some actual
\ha\ emission, although about 75\% of the apparent \hr\ excess can be
accounted for as a consequence of the spectral features of a normal
mid-M dwarf star that is part of the disk population. We calculated an
equivalent width (EW) corresponding to the residual \hr\ excess of
$-0.2$, following the procedure of \citet{Beccari14}, finding a value
of EW(\ha) = 13\,\AA. This is similar to the values they found for
early M stars registering an \ha\ excess in 47~Tuc. 

While the two stars do not have a formal proper motion determination
from KS2, it is clear from direct measurement that they have a similar
and quite substantial proper motion, about 27 mas~yr$^{-1}$. For the
distance range of $500-800$\,pc estimated above, this corresponds to a
transverse velocity of $64 - 100\,\kms$. This is a high velocity range
for the thin disk of the Galaxy, but is consistent with the velocity
range for the thick disk \citep[][p.\,656]{Binney98}.

We can gain further information about CX8 from the combined X-ray
spectrum and flux.  \citet{Pooley02} noted that it has a very soft
spectrum, and suggest a millisecond pulsar nature cannot be excluded.
\citet{Heinke03} suggested that the spectrum showed evidence for an
emission line, and that a foreground AB nature was the most likely
explanation for this system.  We have performed spectral fits to the
combined Chandra data, extracted as described in \citet{Forestell14},
grouping by 15 counts/bin.  We find that fitting power law or hydrogen
neutron star atmospheres \citep[NSATMOS,][]{Heinke06} with the cluster
$N_H$ provides unacceptably poor fits.  Thawing the $N_H$ allows
decent fits, though we then find
$N_H=3.3^{+0.8}_{-0.7}\times10^{21}\,{\rm cm^{-2}}$ for the NSATMOS
fit, which is 10 times higher than the cluster value, which would be
difficult to explain for a millisecond pulsar nature.  A single MEKAL
model \citep{Liedahl95} gives an unacceptable fit, but a double MEKAL
model (as typically found for ABs by e.g.\ \citealt{Dempsey97}) with
the cluster $N_H$ gives acceptable fits ($\chi^2=11.91$ for 9 degrees
of freedom).  The inferred temperatures of $1.15^{+0.33}_{-0.18}$ and
$0.38^{+0.21}_{-0.12}$\,keV, the relative emission measures of the two
components (the higher temperature component having
1.4$^{+1.6}_{-0.7}$ the emission measure of the cooler), and the
inferred luminosity ($L_X(0.5-10\,{\rm keV})=2-5\times10^{29}~\ergs$,
at the distance range estimated above) are all consistent with the
range of BY Dra systems discussed in \citet{Dempsey97}. We also
performed the same fits using unbinned spectra and the C-statistic in
XSPEC, finding results that are consistent within the error bars with
the results above.  Thus, the combined evidence from photometry,
proper motion, and the X-ray spectrum strongly indicates that CX8 is a
pair of foreground, chromospherically active, M-dwarf binaries, likely
of solar or greater abundances.

\newpage
\subsection{Chance Coincidences}

In order to estimate the number of chance coincidences of cluster
stars with \chandra\ source regions, we computed the number of both MS
stars and blue stars expected to fall in each search region. In both
cases we considered stars with $\R > 16$. MS stars were defined by
$\br \ge 0.6$, while blue stars were defined by $\br < 0.6$. The
estimated number of chance coincidences per search area was computed
from the radial surface density for each group of stars times the area
of the search region. No proper-motion cleaning was applied, since in
counting the actual number of objects per search area we did not apply
proper-motion cleaning. The predicted number of MS stars per search
area was typically within about a factor of two of the observed number,
which is consistent given the small-number statistics. On the other
hand, the predicted number of blue stars within each search area was
minuscule, with a median value of about 0.02. The total number of
spurious matches of blue stars with all search areas was predicted to
be 0.9. This indicates that the MS stars that are observed within the
search areas are almost certainly chance superpositions, while the
blue stars are highly likely to be bona fide identifications.

\subsection{Comparison with \citet{Pooley02} and \citet{Thomson12}}

We confirm all of the identifications made by \citet{Pooley02} except
for that of source CX11\@.  \citet{Thomson12} similarly were unable to
confirm this identification.  In this case, the situation is
complicated by two overlapping diffraction rings around nearby bright
stars.  While there is a suggestion of extra flux at the location
indicated by \citet{Pooley02}, it may simply represent the
superposition of the two rings. There are three stars that lie at the
edge of the search area, about 0\farcs3 from the source position. One
of these stars is a subgiant, one is a MSTO star, and the remaining
fainter star has an undetermined \R\ magnitude from KS2\@. Based on
aperture photometry, this star appears likely to lie near the MS\@. As
we discuss in \citet{Forestell14}, source CX11 is within the
positional uncertainty of MSP~D, which shows no evidence of binarity
\citep{Corongiu06}.  Since an isolated neutron star is not expected to
have a detectable optical counterpart, it is most likely that the
coincidence of the three stars with the source CX11 search area is a
chance superposition. In fact, about $1.7 \pm 1.3$ chance
superpositions are expected within the search area for CX11.

\citet{Pooley02} noted a coincidence between source CX17 and a radio
source detected with the Australia Telescope Compact Array. This is
consistent with our classification of the CX17 optical counterpart as
a galaxy, based on the position of CX17 in the center of a diffuse
optical object that clearly seems to be a galaxy (see
Fig.~\ref{f:finding_charts}), interpreting the radio and X-ray
emission as likely both associated with an active galactic nucleus.

We note that \citet{Thomson12} have suggested a different counterpart
to source CX8 than the pair of red stars that we discuss in
\S\ref{source_types} above.  They stated that the $U$ and $B$
photometry of their proposed counterpart indicates that it is a
``faint gap source,'' i.e.\ that it lies between the MS and the
extended blue horizontal branch and thus is a possible CV\@.  However,
from the full WFC3 photometry provided by \citet{Thomson12}, it can be
seen that their suggested counterpart to CX8 lies on the main sequence
in the ($U\!-\!B$, $B$) CMD (the only filters in which they detected
it). Our ACS photometry of this star also indicates that it lies on
the MS in both the (\br, \R) CMD and the (\hr, \R) CMD\@.  Thus, based
on all the available data, we see no evidence for interpreting it as a
CV\@.

\begin{figure*}
\epsscale{0.9}
\plotone{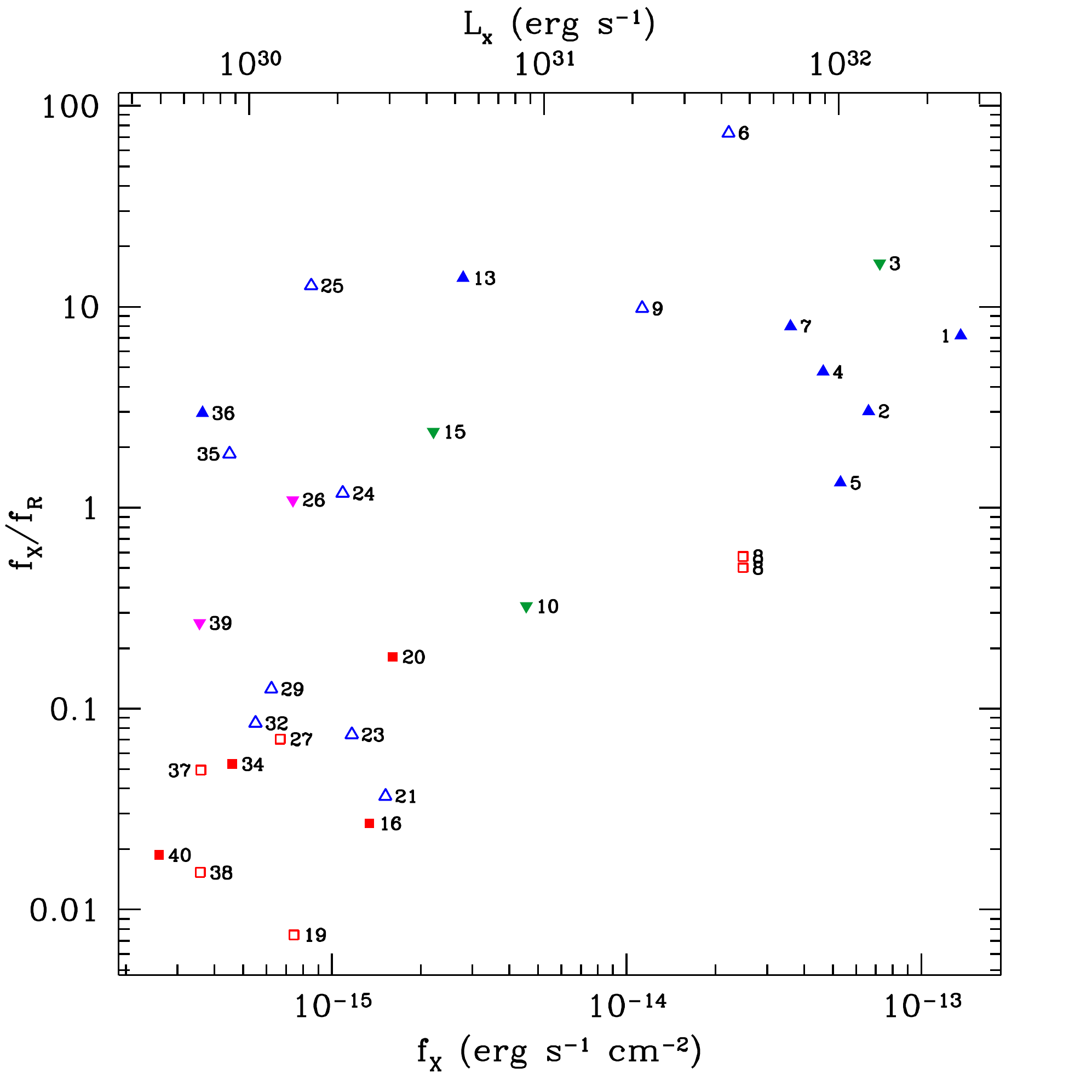}
\figcaption{X-ray to optical \R-band flux ratio vs.\ X-ray flux
  (0.5--8 keV) for CVs (blue triangles), ABs (red squares), and
  galaxies (inverted magenta triangles).  Less certain identifications
  are plotted with open symbols.  The upper axis gives the equivalent
  X-ray luminosity assuming that all objects are at the distance of
  the cluster.  Note that the CVs mostly populate the upper part of
  the diagram, while the ABs mostly populate the lower part.
\label{f:fx_fopt}}
\end{figure*}

We also note that \citet{Thomson12} have pointed out that the
counterpart proposed by \citet{Pooley02} for source CX16, which
shows the photometric characteristics of a BY Draconis star, is
outside of the \citet{Pooley02} error circle.  They suggest another
closer star as a more likely counterpart.  However, we find that the
counterpart proposed by \citet{Pooley02} is the closest star to the
refined source CX16 position from \citet{Forestell14}, falling within
the search area at $1.4~r_{\rm err} = 0\farcs19$ from the X-ray
source location.  Thus, we confirm the \citet{Pooley02} identification
of this source.

\citet{Thomson12} have suggested a possible SX Phoenicis star as a
counterpart to source CX12.\@ However, \citet{Forestell14} were able
to resolve CX12 (with the additional \chandra\ data) into three
sources (CX20, CX23, and CX24).\@ Consequently, the counterpart
suggested by \citet{Thomson12} lies 0\farcs75 from the nearest
X-ray position (CX24), well outside our search radius. We suggest
counterparts to each of CX20, CX23, and CX24.

\subsection{X-ray to Optical Flux Ratios \label{flux_ratio}}

As in \citet{Cohn10}, we have examined the X-ray to optical flux
ratio, $f_X/f_{\rm opt}$, where we take $f_X (0.5\!-\!8\,{\rm keV})$
from \citet{Forestell14} and set $f_{\rm opt} = f_{R_{625}} =
1.07\times10^{-(0.4R_{625}+6)}$.  The latter conversion factor is
computed from the \hst\ flux calibration constants and includes a
small correction for the total extinction of $A_R = 0.10~{\rm mag}$.
The resulting flux ratio is plotted versus $f_X$ in
Fig.~\ref{f:fx_fopt}.  The ratio has been observed to be higher for
accreting sources, such as CVs and LMXBs, than for ABs
\citep[e.g.\ ][]{Edmonds03b,Bassa08}.

We have used this observation to support our identification of the
counterpart to the source CX5 as a possible CV, on the basis of its
high $L_X = 1.1\times10^{32}~\ergs$ and moderately high $f_X/f_{\rm
  opt} = 1.4$.  These values are inconsistent with the other objects
classified as ABs, with the exception of CX8\@.  As noted in
\S\ref{source_types}, our proposed source CX8 counterpart stands out
for its large \ha\ excess, which presumably indicates a high level of
chromospheric activity.  Its high value of $f_X/f_{\rm opt} = 0.5$ is
consistent with this.  As discussed in \S\ref{source_types}, CX8 is
likely to be a foreground object, which would significantly reduce its
inferred $L_X$ value relative to the cluster members.

It can be seen in Fig.~\ref{f:fx_fopt} that the objects classified as
likely or less certain CVs mostly populate the upper part of the
diagram and the likely or less certain AB candidates mostly populate
the lower part.  The median flux ratio is 60 times larger for the CVs
than for the ABs.  The two galaxies for which flux ratios were
calculated fall in between the bulk of the CVs and the bulk of the
ABs.  This suggests that they are normal galaxies, rather than active
galaxies, for which the flux ratio should be at least an order of
magnitude larger.

\begin{figure*}
\centering
\includegraphics*[clip,viewport=18 144 592 718,width=5.5in]{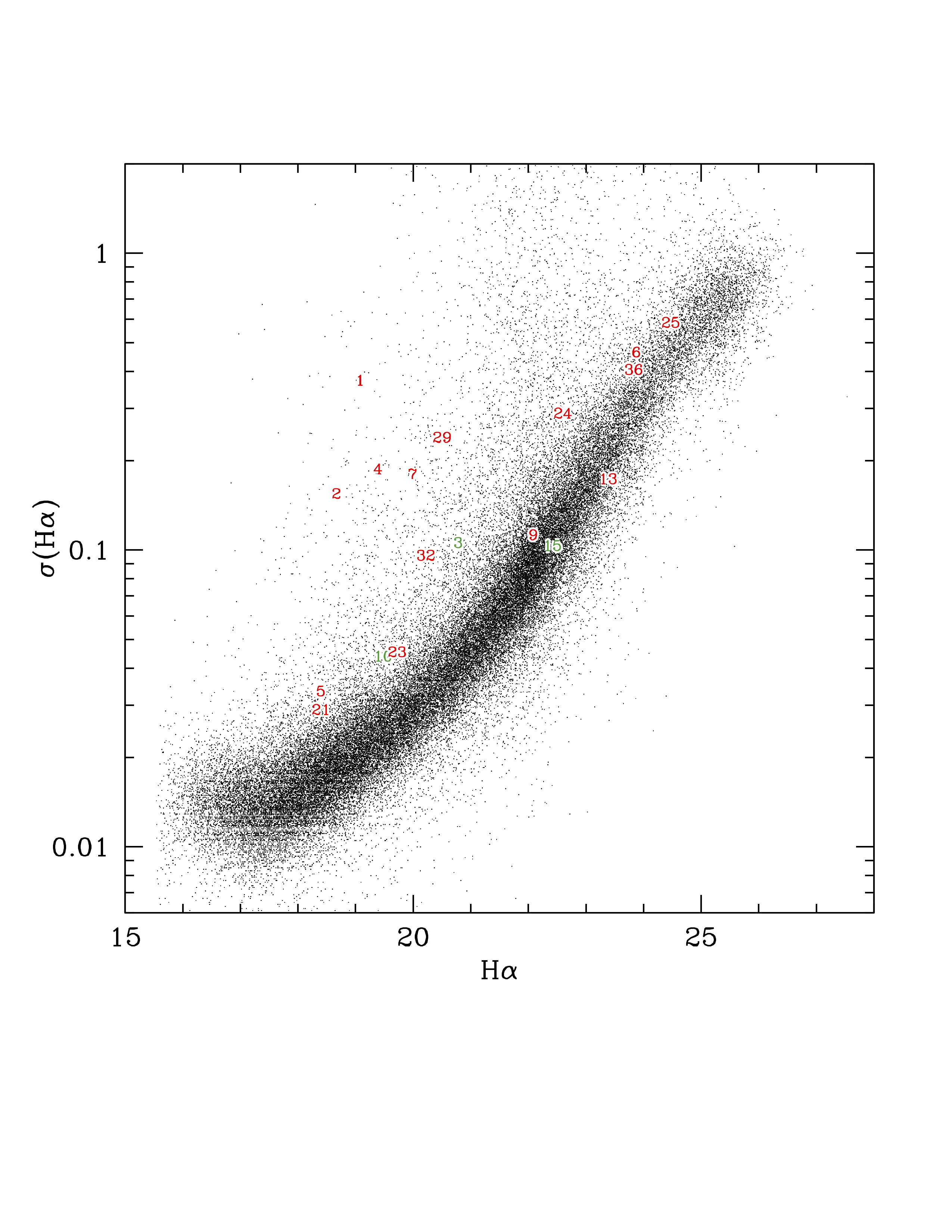}
\figcaption{\ha\ variability versus \ha\ magnitude for candidate CVs.
  The ordinate is the 3$\sigma$-clipped RMS deviation of the up to
  24 \ha\ measurements for each star.  The locations of the CV
  candidates are plotted as numbers over the distribution for all
  stars.  Note that most of the bright CVs show a variability signal,
  i.e.\ they lie above the relation defined by the majority of the
  stars.  
\label{f:variability_CV}}
\end{figure*}

\begin{figure*}
\centering
\includegraphics*[clip,viewport=18 144 592 718,width=5.5in]{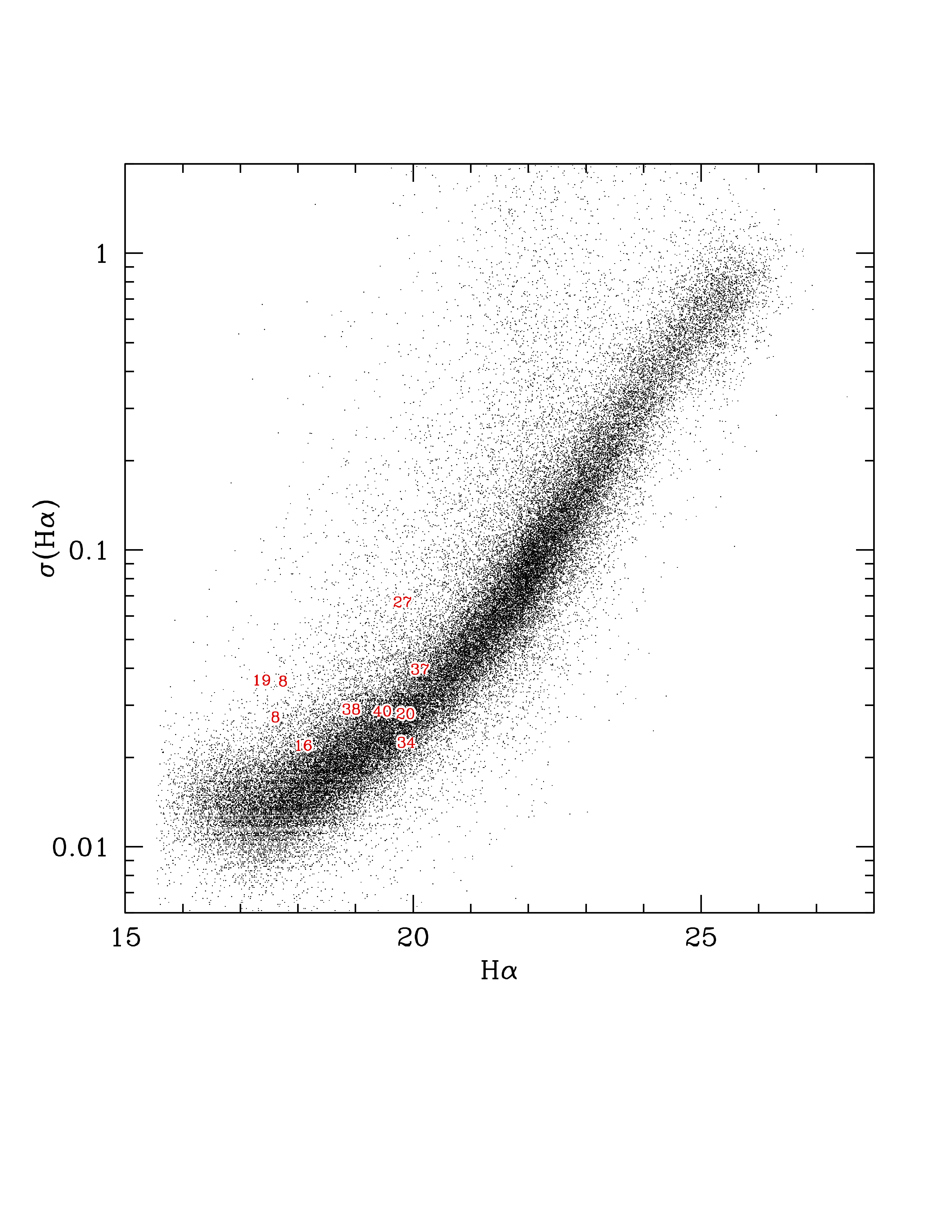}
\figcaption{\ha\ variability versus \ha\ magnitude for candidate ABs.
  Note that the ABs, on average, show lower variability than the CVs
  illustrated in Fig.~\ref{f:variability_CV}.
\label{f:variability_AB}}
\end{figure*}

\subsection{Variability} \label{variability}

Since our dataset provides a 24-exposure time sequence of
\ha\ exposures, with two exposures per orbit, it was possible to
investigate optical variability of the detected objects.  The time
sequence samples time scales shorter than about one hour (the
visibility period per \hst\ orbit) and also time scales of days to
months.  As a measure of the variability, we first adopted the
3$\sigma$-clipped RMS deviation, which we used in our study of
NGC~6397 \citep{Cohn10}.\footnote{Following a suggestion of the
  referee, we now compute the fractional RMS deviation of the fluxes
  and express this in magnitude units.}  Since the use of sigma
clipping rejects outliers, this approach results in a measure of
variability that is most sensitive to the orbital variability of
binary systems, rather than large-amplitude fluctuations of CVs.

We plot $\sigma$(\ha) versus mean \ha\ magnitude in
Figs.~\ref{f:variability_CV} and \ref{f:variability_AB}.  It can be
seen in these figures that most stars fall along a ``fundamental''
sequence of increasing $\sigma$(\ha) with increasing magnitude.  We
interpret stars falling above this sequence as showing evidence of
optical variability, although some degree of photometric scatter is
also clearly present.  We have plotted the locations of the
\chandra\ source counterparts on this diagram.  A number of the
counterparts show apparently significant variability.  This group
includes many of the bright CV counterparts, viz.\ sources CX1, CX2,
CX4, CX7, CX29, and CX32 as seen in Fig.~\ref{f:variability_CV}.  Four
of the remaining bright CV candidate counterparts, sources CX5, CX21,
and CX23, show marginal evidence of variability, falling just above
the fundamental sequence.  Of the faint CV candidate counterparts,
sources CX6 and CX24 show marginal evidence of variability and sources
CX9, CX13, and CX36 fall on the high side of the fundamental
sequence. At the magnitude of these faint CVs, $\ha \approx 24$, the
large typical photometric uncertainty of about 0.3 mag makes it
difficult to detect moderate variability.  The amplitude of the
variability measured by $\sigma$(\ha) for the bright CVs is about 0.1
-- 0.3 mag, which is typical for orbital variations.  The variability
plot for the ABs is shown in Fig.~\ref{f:variability_AB}.  It can be
seen from this figure that sources CX8, CX19, and CX27 show evidence
of variability.  As discussed in \S\ref{source_types}, source CX8
appears to be a foreground system.

\begin{figure*}
\centering
\includegraphics*[clip,viewport=18 144 592 718,width=5.5in]{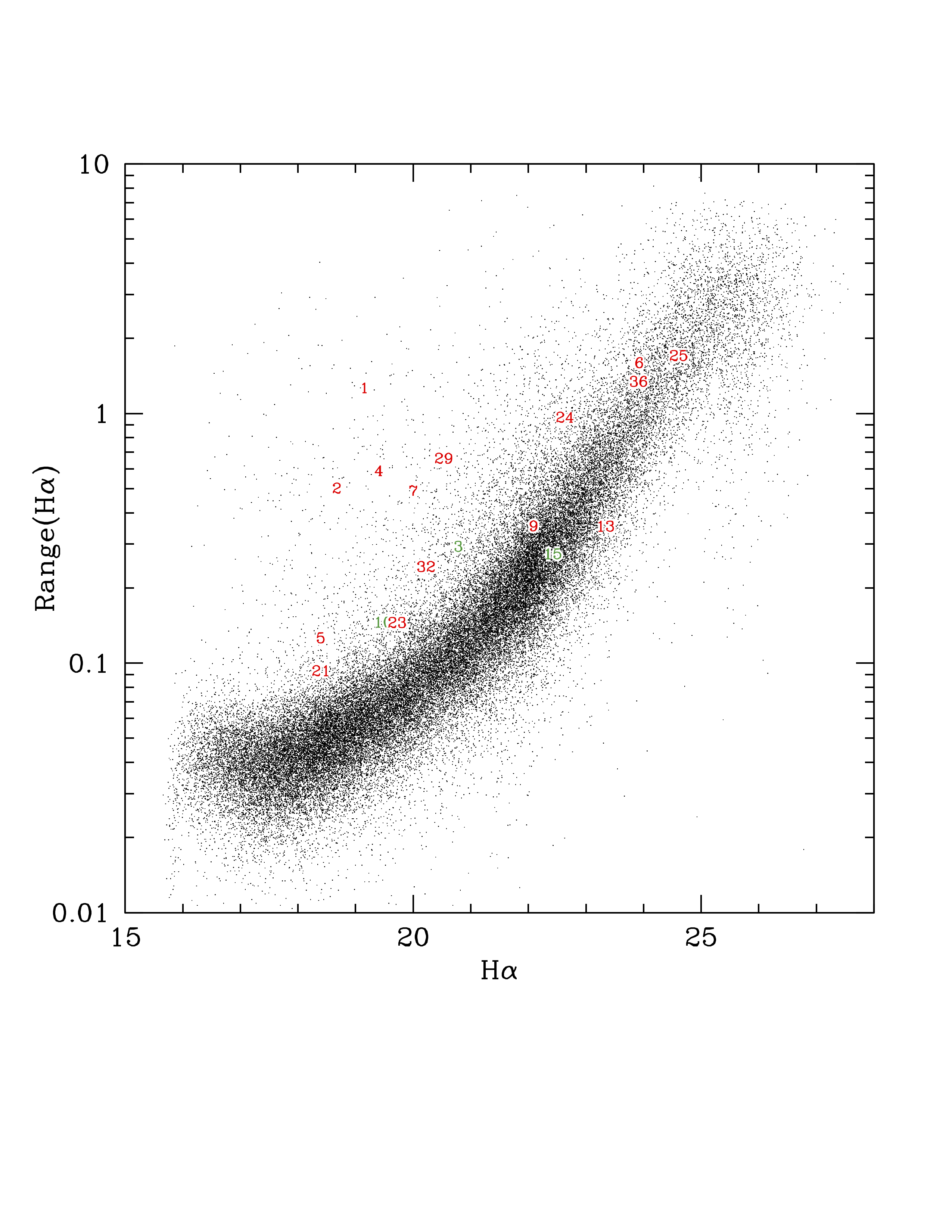}
\figcaption{\ha\ range versus \ha\ magnitude for candidate CVs.  The
  ordinate is the full range of the up to 12 \ha\ measurements for
  each star, as we took the faintest of each pair of \ha\ measurements
  as representing the best estimate of the \ha\ magnitude for that
  orbit.  The CV candidates are plotted as numbers over the
  distribution for all stars. Note the similarity between this figure
  and Fig.~\ref{f:variability_CV}, other than the greater amplitude of
  the variability here. 
\label{f:variability_range}}
\end{figure*}

We also investigated the total range of the \ha\ magnitudes for each
object in order to search for outburst behavior in the CV
counterparts. In order to filter out cosmic-ray events that affect
some of the individual magnitude measurements, we chose the fainter of
the magnitude measurements from each of the pair of \ha\ frames per
orbit. We discarded orbits for which there was only one successful
\ha\ magnitude measurement. We plot the total range of the orbital
\ha\ magnitudes versus \ha\ magnitude for all the stars, with the CV
counterparts indicated, in Fig.~\ref{f:variability_range}.  Five of
the bright CV counterparts have a variation range of $\sim 0.5 -
1.0$~mag. The source CX1 stands out with the largest range of \ha,
with a value of $\sim 1.3$~mag.  While some of this variation range
may be due to photometric uncertainty, examination of the individual
images indicates that it is largely due to actual flux variation.
Thus, source CX1 appears to have undergone a dwarf nova outburst
during the approximately 180\,d interval over which the data were
obtained.  We note that \citet{Kaluzny09} observed a 1.5 mag amplitude
outburst from the CX4 counterpart, in a study of stellar variability
in NGC~6752 that included the locations of all of the \citet{Pooley02}
\chandra\ sources.  \citet{Thomson12} observed an outburst of 1.5 mag
amplitude from the CX1 counterpart and an outburst of 6 mag amplitude
from the CX7 counterpart.

\citet{Kaluzny09} detected periodic variability from a suggested
counterpart to CX19.\@ They found a period of 0.11306\,d, with an
amplitude of a few hundredths of a mag.  They note that if this
modulation is due to ellipsoidal variations, then the orbital period
would be twice this large, viz.\ 0.226\,d.  While we were not able to
confirm the periodicity of this counterpart with a period-folding
analysis of the \ha\ time series, Fig.~\ref{f:variability_AB}
indicates that the object is variable, with an amplitude of about 0.04
mag.  We note that \citet{Thomson12} did not find any evidence of the
\citet{Kaluzny09} periodicity in their WFC3 near-UV data.
\citet{Kaluzny09} suggest that the counterpart to CX19 is a close
binary hosting a neutron star or a black hole.  However, its location
to the red of the MS in the (\br, \R) CMD and its low X-ray to optical
flux ratio of $f_X/f_{\rm opt} = 0.007$ suggest instead that it is
likely to be an AB\@. In this case, the longer orbital period of
0.226\,d is strongly preferred, since even the contact binaries in
47~Tuc in this range of stellar masses have periods of at least
0.20\,d \citep{Albrow01}.

\subsection{Spatial Distribution}

We determined the cluster center by iterative centroiding in a
12\arcsec\ radius aperture using a sample of stars with magnitudes in
the range $\R < 20.5$, which extends to 4 mag below the MSTO\@.  The
resulting center of $\alpha = 19^{\rm h}~10^{\rm m}~52\fs12,~
\delta = -59\arcdeg~59\arcmin~4\farcs4$ agrees well with the recent center
determinations by \citet{Goldsbury10} and \citet{Thomson12}, differing
from the former by 0\farcs11 and from the latter by 0\farcs15.
Experimentation with the centroiding aperture size and the stellar
sample definition indicates that the center position is uncertain by
about 0\farcs5.

\begin{figure*}
\plottwo{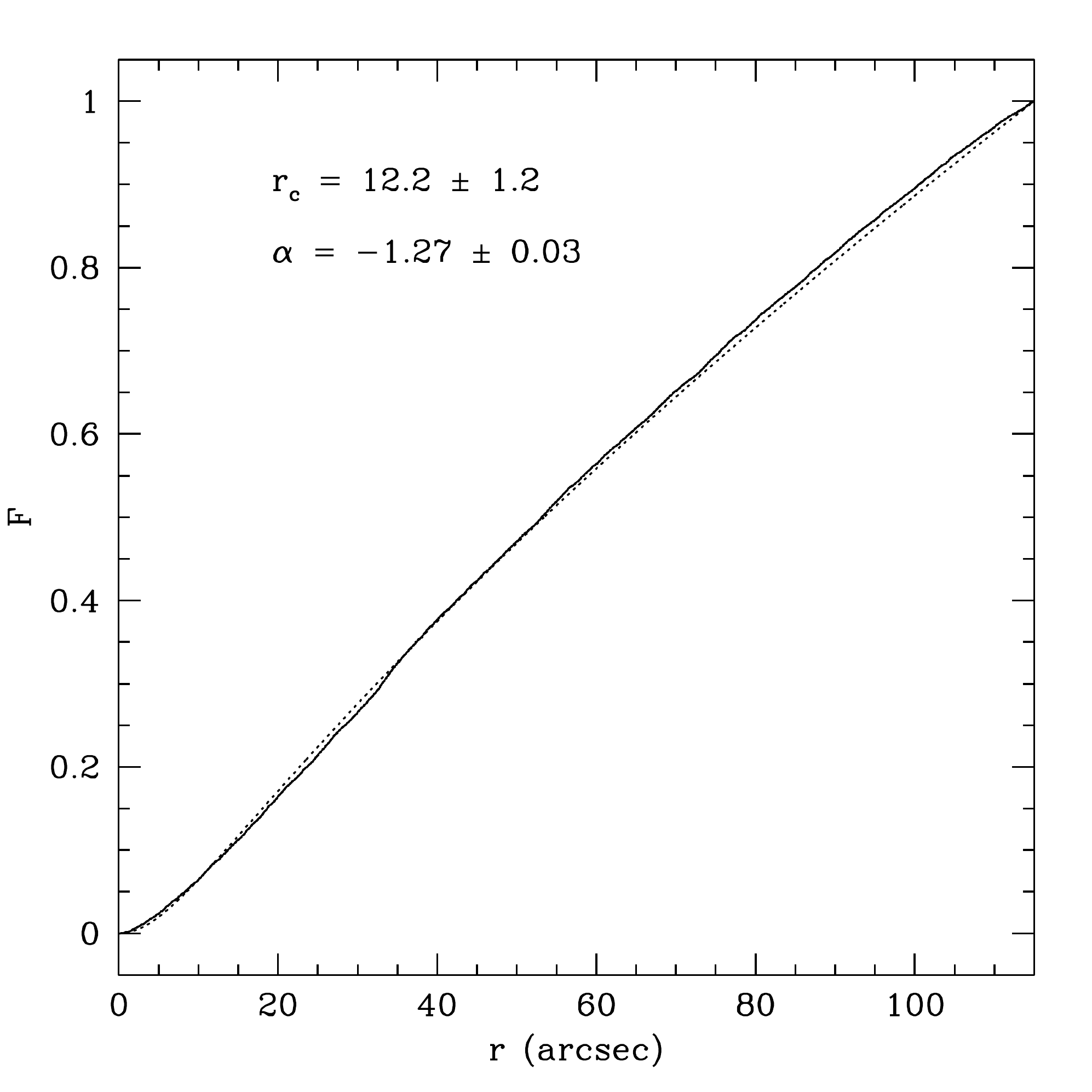}{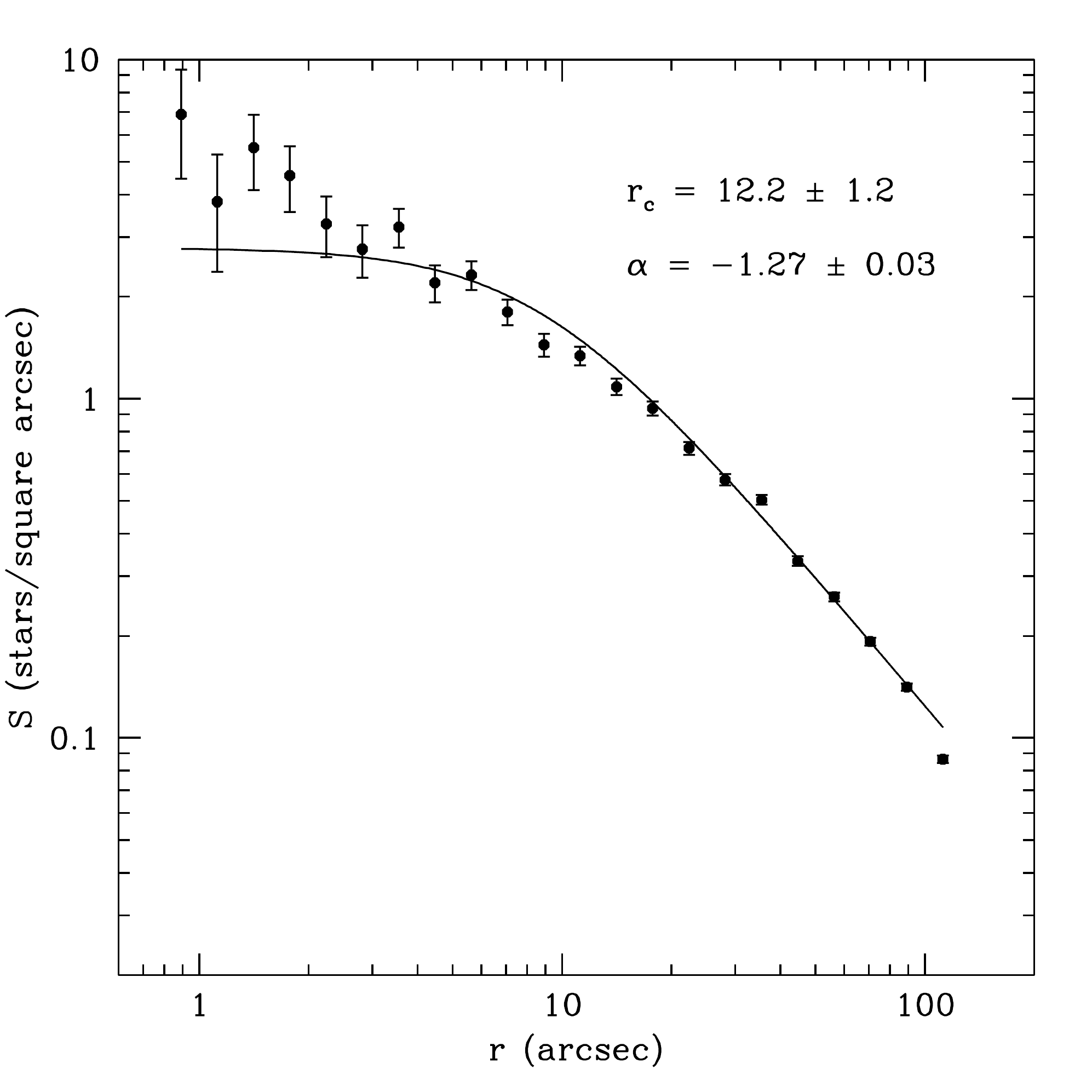}
\figcaption{Radial surface-density profile for the MSTO group.  Left
  panel: cumulative radial distribution (solid line) with a
  cored-power-law fit to 115\arcsec\ (dashed line).  Right panel:
  binned surface-density profile with the same cored-power-law fit.  A
  one-sample K-S test indicates that the data are consistent with the
  fit with a probability of 14\%.  Nevertheless, the curve falls
  systematically below the data for $r<5\arcsec$ and above the data
  for $5\arcsec < r < 20\arcsec$.
\label{f:radial_profile_1}}
\end{figure*}

\begin{figure*}
\plottwo{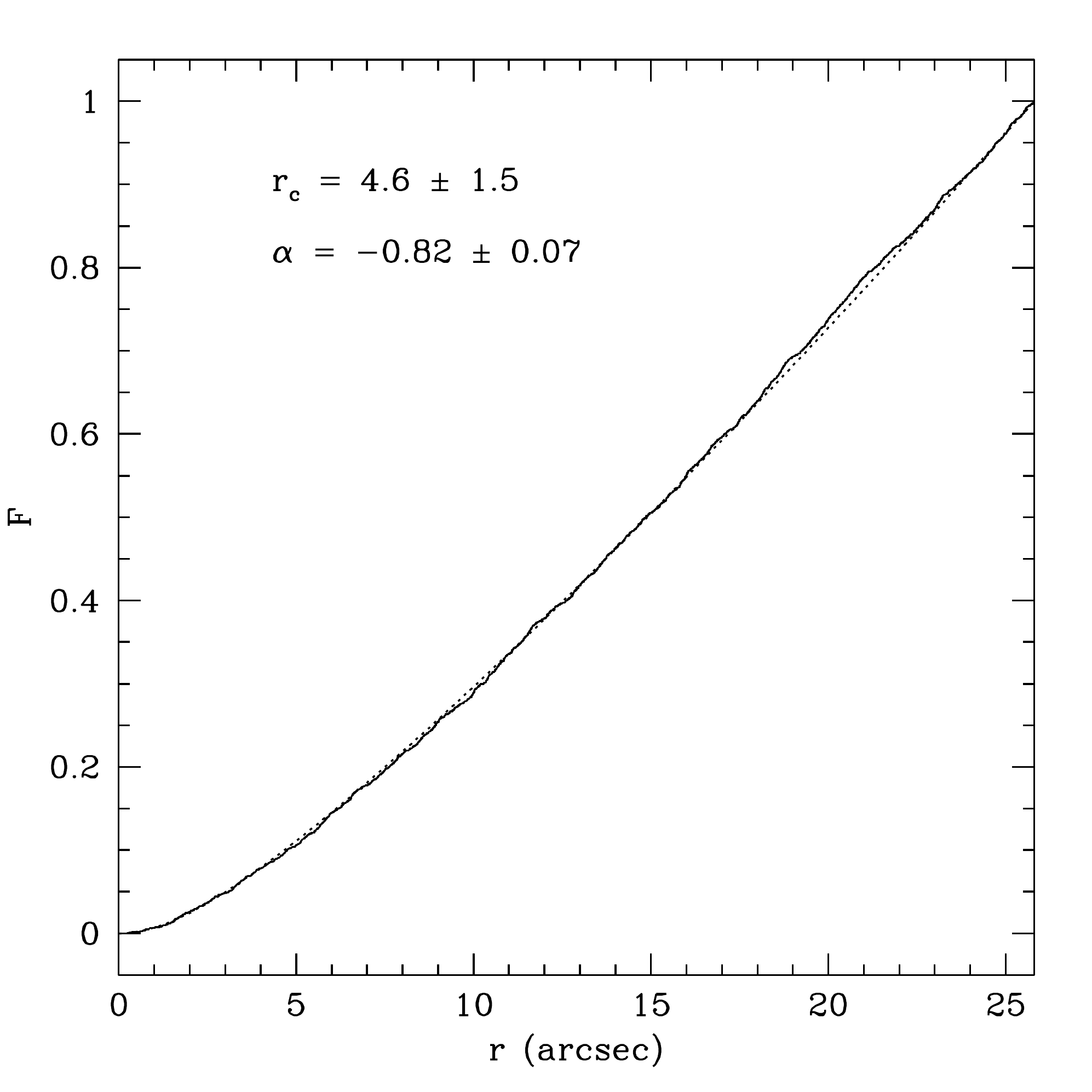}{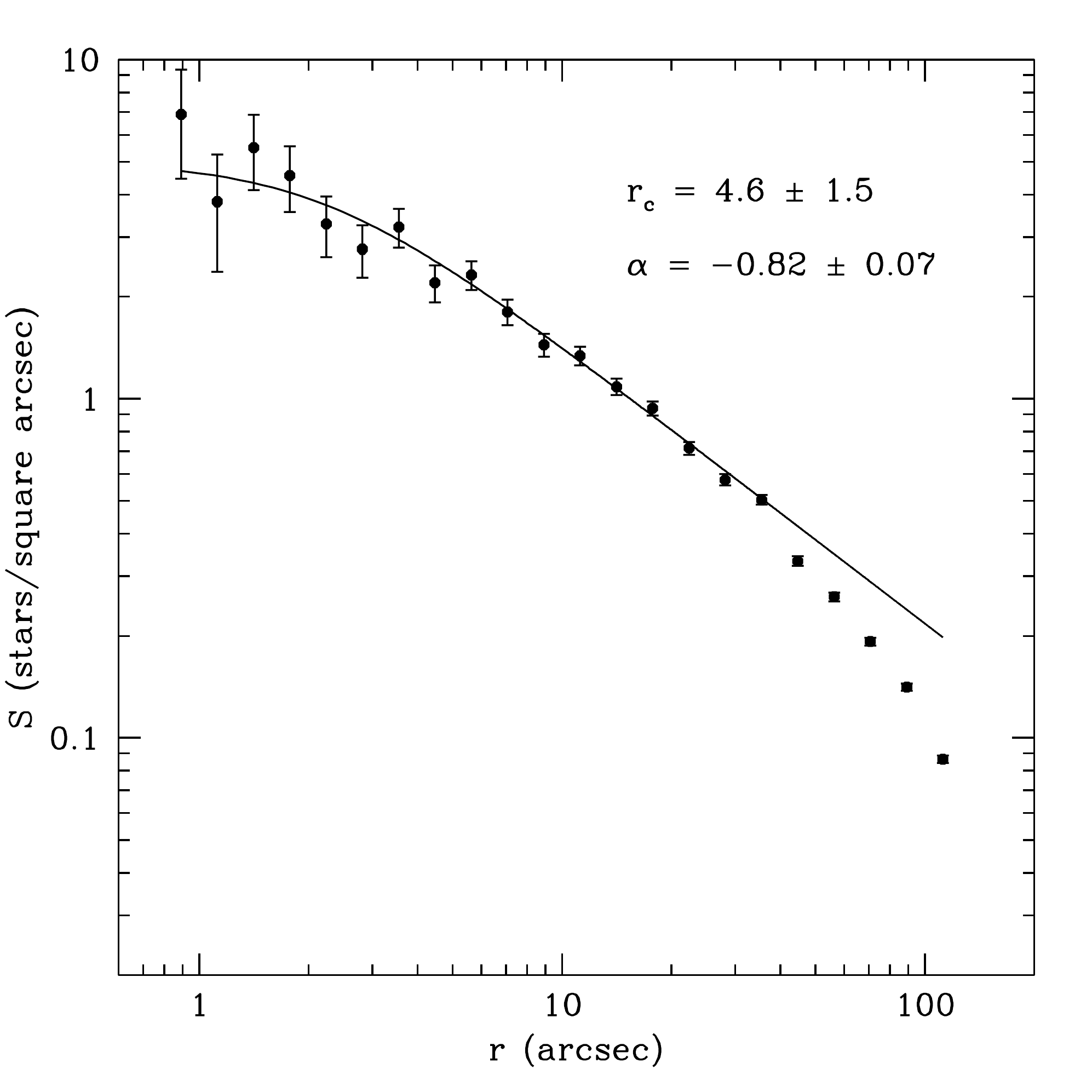}
\figcaption{Radial surface-density profile for the MSTO group as in
  Fig.~\ref{f:radial_profile_1}.  Left panel: cumulative radial
  distribution (solid line) with a cored-power-law fit to
  25\farcs8 = 0.5 pc (dashed line).  Right panel: binned
  surface-density profile with the same cored-power-law fit.  Note that
  the fit to the inner profile is much better than that shown in
  Fig.~\ref{f:radial_profile_1}.
\label{f:radial_profile_2}}
\end{figure*}

We next determined both the cumulative radial distribution and a
binned radial profile for a MSTO group of stars with magnitudes in
the range $16 \le \R < 18.5$, which extends to 2 mag below the
MSTO\@.  These are shown in Fig.~\ref{f:radial_profile_1}.  In order
to assess the overall behavior of the cluster profile, we fitted the
cumulative radial distribution of the MSTO group with a ``generalized
King model,'' which we have also called a ``cored power law.''  As
discussed by \citet{Cohn10}, this takes the form
\begin{equation}
\label{eqn:Cored_PL} 
S(r) = S_0 \left[1 + \left({r \over r_0}\right)^2 \right]^{\alpha/2},
\end{equation}
with the core radius $r_c$ related to the scale parameter $r_0$ by,
\begin{equation}
r_c = \left(2^{-2/\alpha} -1 \right)^{1/2} r_0\,.
\end{equation}
Fig.~\ref{f:radial_profile_1} shows the resulting maximum-likelihood
fit where the data have been fitted to a limiting radius of $r_h =
115\arcsec$.  As seen in Fig.~\ref{f:mosaic}, the region within $r_h$
is fully covered by the ACS/WFC imaging.  While the model provides a
statistically acceptable fit to the profile over this data range, it
can be seen in the right panel of Fig.~\ref{f:radial_profile_1} that
the fit is systematically low in the central region of the cluster and
high in an intermediate radial range.  Moreover, the best-fit slope of
$\alpha = -1.28$ also does not agree with that expected for an
``analytic King model,'' for which $\alpha = -2$.  Thus, the radial
profile of NGC~6752 is not well fitted by a single-mass King model,
indicating that it does not have a normal-core structure like clusters
such as 47~Tuc \citep{Howell00}.  This suggests that NGC~6752 is
in a post-collapse state of evolution as concluded by several previous
studies noted above \citep{Djorgovski86,Ferraro03a,Thomson12}.  In
order to evaluate the parameters of the expected post-collapse
surface-density cusp, we fitted the cored-power-law model
(Eqn.~\ref{eqn:Cored_PL}) to the cumulative radial distribution using
an outer limiting radius of 25\farcs8, corresponding to a radial
scale of 0.5~pc, which is the limiting radius adopted by
\citet{Lugger95} for the cored-power-law fits they presented for 15
candidate core-collapsed clusters.  The fits are shown in
Fig.~\ref{f:radial_profile_2}.  The best-fit parameter values are $r_c
= 4.6'' \pm 1.5''$ and $\alpha = -0.82 \pm 0.07$. These indicate that
NGC~6752 has a well-resolved core, with a surrounding cusp slope that
agrees well with the mean core-collapse slope of $-0.84 \pm 0.10$
found by \citet{Lugger95}.  We note that they similarly found that a
cored power law gave a good fit to the central $U$-band
surface-brightness profile of NGC~6752 with best-fit parameters of
$r_c = 6.7'' \pm 1.9''$ and $\alpha = -0.97 \pm 0.15$.  These two sets
of best-fit parameter values are consistent with each other to within
1$\sigma$.  In retrospect, we conclude that our ``conservative''
interpretation, in \citet{Lugger95}, that NGC~6752 is not required to
be in a post-collapse state, is too conservative and that the
surface-density profile of NGC~6752 provides good evidence that the
cluster has indeed experienced core collapse.

\begin{figure}
\epsscale{1.2}
\plotone{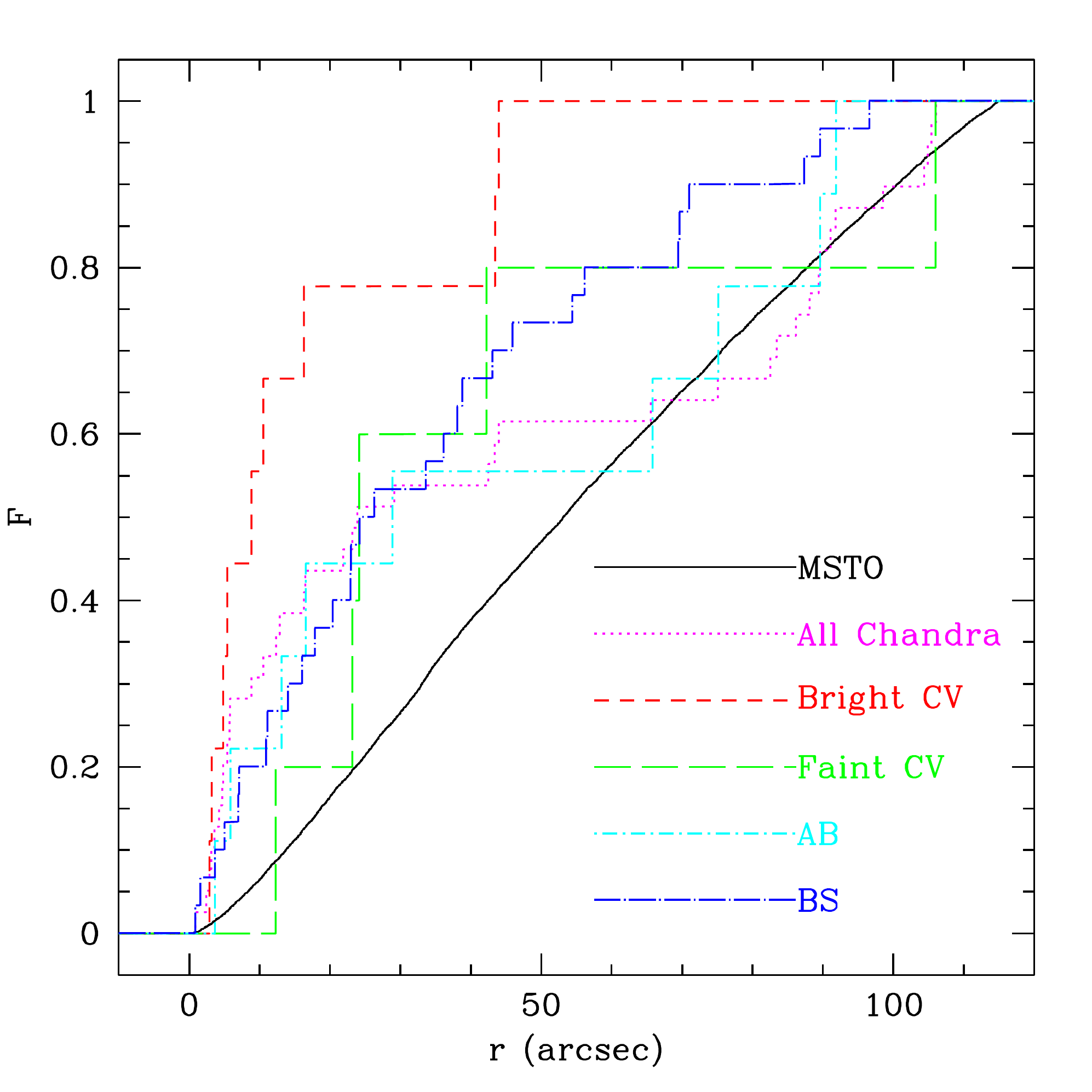}
\figcaption{Cumulative radial distributions for selected stellar
  groups.  Note that the \chandra\ sources, the bright CVs, and the
  BSs show significant central concentration ($p \lesssim 1\%$)
  relative to the MSTO group.  Fitting information and K-S sample
  comparisons for these stellar groups are given in
  Table~\ref{t:Cored_PL_fits}.  
\label{f:cumulative_radial_distributions}}
\end{figure}

\begin{figure}
\epsscale{1.2}
\plotone{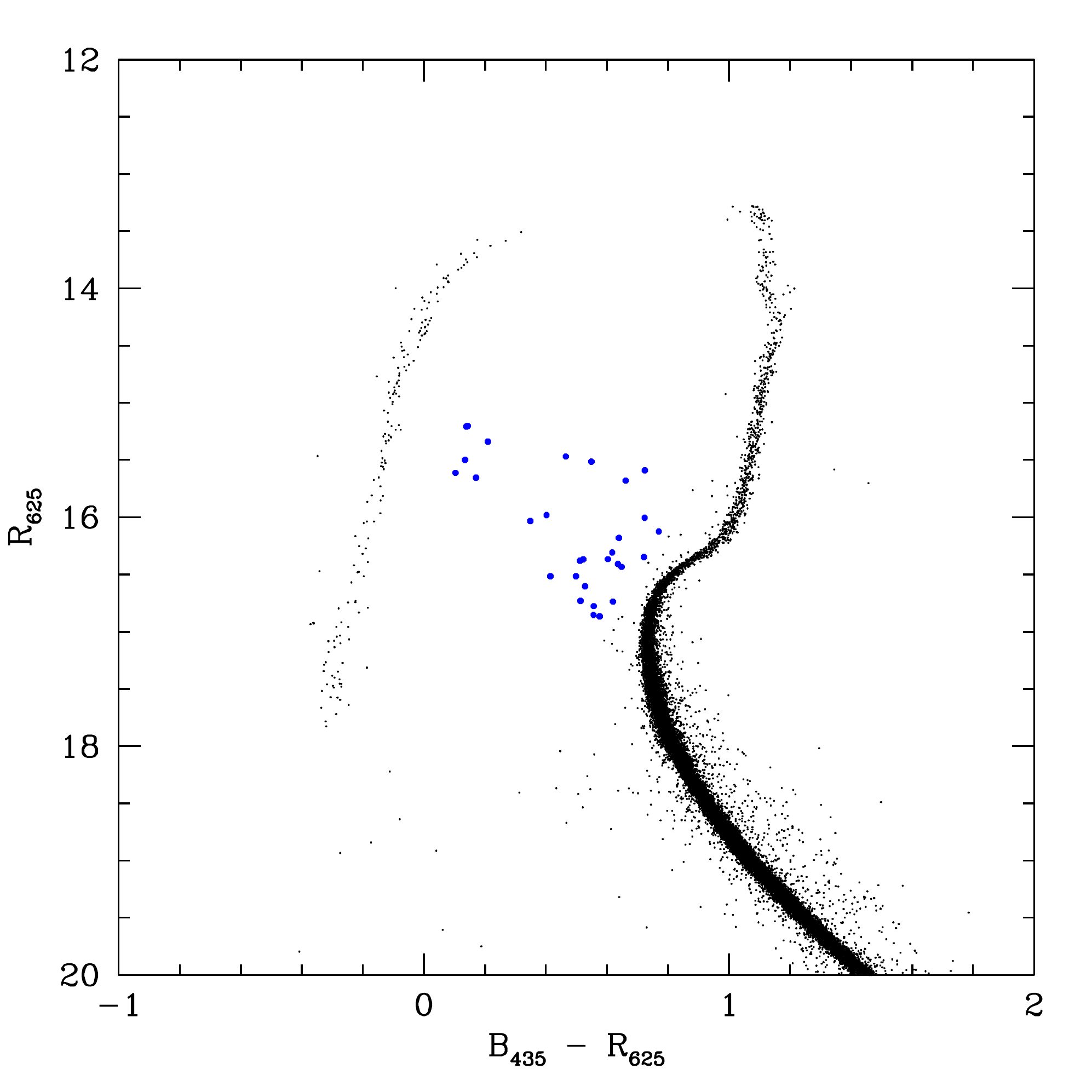}
\figcaption{Color-magnitude diagram showing the selection of blue
  stragglers for the radial distribution analysis.  A conservative
  selection criterion was used to choose 30 BS candidates within the
  half-mass radius.  
\label{f:CMD_BS}}
\end{figure}

We next examined the radial profiles of a number of different stellar
groups by comparing the cumulative radial distributions shown in
Fig.~\ref{f:cumulative_radial_distributions}.  The particular groups
considered are the MSTO group described above, all of the 39
\chandra\ sources within $r_h$, the nine brightest CVs, the five
faintest CVs, the ABs, and a blue straggler (BS) group selected from
the CMD as illustrated in Fig.~\ref{f:CMD_BS}.  As can be seen in
Fig.~\ref{f:cumulative_radial_distributions}, the \chandra\ sources,
the bright CVs, and the BSs show strong central concentration relative
to the MSTO group.  In order to quantify the significance of the
differences in the distributions, we performed Kolmogorov-Smirnov
(K-S) comparisons of each sample with the MSTO group.  The results are
given in Table~\ref{t:Cored_PL_fits}, where the probability, $p$, of
the two samples being drawn from the same parent distribution is
listed.  The distributions of the \chandra\ sources, the bright CVs,
and the BSs differ very significantly ($p<1\%$) from that of the MSTO
group. The faint CV and AB distributions do not differ from the MSTO
group at a significant level ($p=31\%$ and $p=27\%$, respectively). A
direct comparison of the bright and faint CVs indicates that these two
groups differ at the 6\% level. While this misses a 5\% cutoff for
statistical significance, it clearly suggests a meaningful difference
between the two distributions.

\begin{deluxetable*}{lrcccccr}
\tabletypesize{\normalsize}
\tablecolumns{8}
\tablewidth{0pt}
\tablecaption{\textbf{Cored-power-law Model Fits to 115\arcsec} 
\label{t:Cored_PL_fits}}
\tablehead{
\colhead{Sample} & 
\colhead{$N$\tablenotemark{a}} &
\colhead{$q$} & 
\colhead{$r_c~(\arcsec)$} &
\colhead{$\alpha$} &
\colhead{$m~(\msun)$} &
\colhead{$\sigma$\tablenotemark{b}} &
\colhead{K-S prob\tablenotemark{c}}
} 
\startdata
MSTO       &10016 &   1.0            & $12.2 \pm 1.2$ & $-1.27 \pm 0.03$ & $0.80 \pm 0.05$ & \nodata & \nodata \\
\chandra\ sources
           &   39 &  $1.33 \pm 0.12$ & $ 8.5 \pm 0.8$ & $-2.07 \pm 0.28$ & $1.06 \pm 0.10$ & 2.4     & 0.080\%  \\
bright CV  &    9 &  $2.03 \pm 0.35$ & $ 6.0 \pm 0.8$ & $-3.60 \pm 0.79$ & $1.62 \pm 0.28$ & 2.9     & 0.042\% \\
faint CV   &    5 &  $1.27 \pm 0.24$ & $ 9.0 \pm 2.0$ & $-1.88 \pm 0.54$ & $1.02 \pm 0.19$ & 1.1     & 31\%    \\
AB         &    9 &  $1.31 \pm 0.24$ & $ 8.8 \pm 1.8$ & $-1.97 \pm 0.54$ & $1.05 \pm 0.19$ & 1.3     & 27\%    \\
BS         &   30 &  $1.41 \pm 0.11$ & $ 8.1 \pm 0.7$ & $-2.21 \pm 0.26$ & $1.13 \pm 0.09$ & 3.2     & 0.54\%  \\
\enddata
%
\tablenotetext{a}{Size of sample within 115\arcsec\ of cluster center}
\tablenotetext{b}{Significance of mass excess above MSTO mass}
\tablenotetext{c}{K-S probability of consistency with MSTO group}
\end{deluxetable*}

In order to further investigate the spatial distribution of the
various \chandra\ sources and BSs in NGC~6752, we carried out
maximum-likelihood fits of the cored-power-law model to the
surface-density distributions of the groups shown in
Fig.~\ref{f:cumulative_radial_distributions}.  As discussed by
\citet{Cohn10}, this procedure provides an estimate for the
characteristic mass of an object in each group relative to the MSTO
mass.  As in \citet{Cohn10}, we adopt the approximation that the mass
groups above the MSTO mass are in thermal equilibrium.  In this case,
the surface-density profile for a mass group with mass $m$ is given by
Eqn.~\ref{eqn:Cored_PL} with a slope parameter $\alpha$ related to the
turnoff-mass slope $\alpha_{\rm to}$ by
\begin{equation}
\alpha = q\, (\alpha_{\rm to} - 1) + 1
\end{equation}
where $q = m/m_{\rm to}$.

Table~\ref{t:Cored_PL_fits} gives the results of maximum-likelihood
fits of Eqn.~\ref{eqn:Cored_PL} to the turnoff-mass stars,
\chandra\ sources, CVs, ABs, and BSs.  As can be seen from the table,
the $q$ values for all of the groups exceed unity, indicating that the
characteristic masses exceed the turnoff mass.  We assume a MSTO mass
of $0.80 \pm 0.05\,\msun$, based on the study of \citet{Gruyters14},
who found a MSTO mass of 0.79\,\msun. For the \chandra\ sources, bright
CVs, and BSs, the excesses are significant at the $2.4\sigma-3.2\sigma$
level.  We note that the results of this analysis are supported by the
K-S comparison results given in the last column of the table.

The inferred mass range for the bright CVs ($1.6 \pm 0.3\,\msun$) is
similar to what we found for NGC~6397\@.  The median \R-band absolute
magnitude for these systems is about $M_R \approx 7$. Assuming that
the \R-band flux is dominated by the secondary, this implies a
secondary mass of about 0.6\,\msun, based on the isochrones of
\citet{Baraffe97}. The corresponding WD mass is $M_{\rm WD} \sim
1.0\,\msun$, which is consistent with the value of $0.83 \pm
0.23\,\msun$ found by \citet{Zorotovic11}.  The inferred mass range
for the faint CVs ($1.0\pm 0.2\,\msun$) is consistent with a somewhat
lower mass white dwarf (e.g.\ $M_{\rm WD} \sim 0.8\,\msun$) and a
secondary mass that has been whittled down to $\lesssim 0.2\,\msun$.
In such a system, the optical flux would be dominated by the white
dwarf, as observed here.

\newpage
\section{Discussion \label{discussion}}

As in NGC~6397, the bright CVs have a more centrally concentrated
distribution than the faint CVs.  As we discussed for that cluster,
this is consistent with the bright CVs representing a recently formed
population that is produced by dynamical interactions near the cluster
center.  These interactions may likely include exchange interactions
in which a heavy white dwarf displaces a member of a primordial
binary.  As these CVs age, the secondary loses mass and the accretion
rate drops, leading to a reduction in both the optical and X-ray
luminosity \citep{Howell01}. At the same time, two-body interactions
with singles and other binaries scatter the CVs into larger orbits
that put them at increasing distance from the cluster center.

We note that \citet{Hong17} have recently reported Monte-Carlo
simulations of globular cluster dynamical evolution that include CV
formation from primordial binaries and subsequent evolution. They find
that the CVs are more centrally concentrated than the MSTO-mass stars,
with the effect being stronger for the CVs that form from primordial
binaries that have undergone an exchange encounter. This exchange
group includes a population of CVs that are more massive than those
formed from primordial binaries that have not undergone an exchange
encounter. The resulting greater central concentration of the more
massive CVs is qualitatively consistent with our inference that the
brighter CVs in NGC 6752 are more massive than the fainter
ones. \citet{Hong17} note that the radial distribution of the CVs
reflects, ``the remaining memory of the CV formation history and
progenitor masses,'' as well as the present CV mass. Thus, our estimate
of CV masses based on the assumption that they have achieved their
equilibrium radial distribution should be viewed as a first
approximation to a complex process. 

\begin{figure*}
\epsscale{1.1}
\plotone{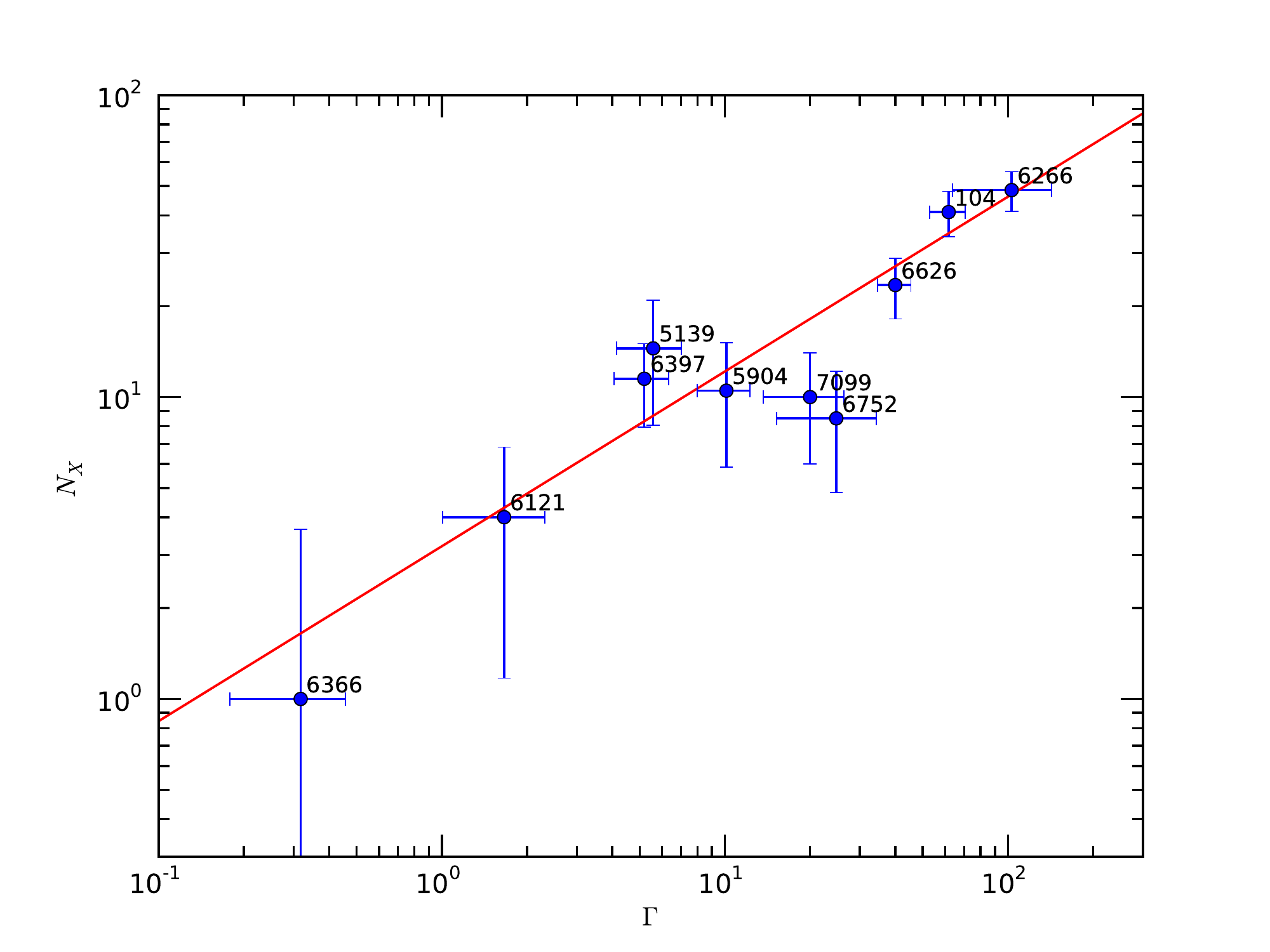}
\figcaption{Number of X-ray sources versus the stellar encounter rate
  $\Gamma$ based on data from \citet{Bahramian13}.  The red line shows
  a linear regression in log-log space, corresponding to the relation
  $N_X \propto \Gamma^{0.58\,\pm0.10}$.
\label{f:Gamma_Nx}}
\end{figure*}

\newpage
\subsection{Stellar Encounter Rate}

\citet{Bahramian13} have reexamined the relationship between the
number of X-ray sources in a cluster and the stellar encounter rate
that was originally found by \citet{Pooley03}.  Fig.~\ref{f:Gamma_Nx}
shows the \citet{Bahramian13} dataset. The errors in the encounter
rate $\Gamma$ are adopted from their study, while the errors in the
background-corrected source counts include the Poisson errors in both
the total source counts and the background counts.  The line in
Fig.~\ref{f:Gamma_Nx} is a linear regression that corresponds to the
relation $N_X \propto \Gamma^{0.58\,\pm\,0.10}$. The slope value is
consistent with the value of $0.74\pm0.36$ obtained by
\citet{Pooley03}. It can be seen from Fig.~\ref{f:Gamma_Nx} that the
core-collapsed clusters NGC~6752 and NGC~7099 both fall below the mean
relation defined by the entire cluster sample.  NGC~6752 is
3.0$\sigma$ below the relation, while NGC~7099 is 2.0$\sigma$
below.  In contrast, \citet{Pooley03} found that the core-collapsed
cluster NGC~6397 lies above the relation.  This deviation of NGC~6397
from the relation is not statistically significant in our
analysis. Our linear regression analysis quantifies the suggestion of
\citet{Bahramian13} that core-collapsed clusters underproduce X-ray
binaries relative to their computed encounter rates. This may indicate
that binary destruction/ejection is a more vigorous process in
core-collapsed clusters than in those that are in a pre-collapse
state.

\citet{Ivanova06} found, from their Monte-Carlo simulations of CV
formation and evolution in globular cluster environments, that
clusters that underwent core collapse in the past 1 -- 2 Gyr will have
a depleted number of CVs, since CVs that have been destroyed during
core collapse will not yet have been replaced by newly formed
CVs. However, the recent Monte-Carlo simulations of CV evolution in
dynamically evolving clusters by \citet{Belloni17} do not show such a
clear effect.

\section{Summary \label{summary}}

We have searched for optical counterparts to the 39 \chandra\ sources
that lie within the half-mass radius of NGC~6752, using \hst\ ACS/WFC
imaging in \B, \R, and \ha.  Based primarily on CMD classification, we
found plausible counterparts to 31 of the sources.  These include 16
likely or less certain CVs, nine likely or less certain ABs, three
galaxies, and three more likely than not AGNs\@.  Our CV/AB
discrimination that is based on CMD location is generally supported by
the X-ray to optical flux ratios for these identifications.  The CVs
have, in most cases, significantly higher values of $f_X/f_{\rm opt}$
than do the ABs, as expected.

In comparison to our results for NGC~6397, where all of the CV
candidates exhibited a strong \ha\ excess in the (\hr, \R) CMD
\citep{Cohn10}, many of the CV candidates reported here do not. We
examined the color-color diagram to further investigate this issue and
found that all but three of our CV candidates do show a \ha\ excess
relative to other stars of the same \br\ color. However, it is not
clear why the \ha\ excesses of the CV candidates in NGC~6752 are
generally weaker than those of the CV candidates in NGC~6397. 

As expected, most of the bright CV candidates registered significant
time variability. The amplitude of the variability, as measured by the
3$\sigma$-clipped RMS of the \ha\ time series, is typically $\sim
0.1-0.3$~mag, which is characteristic of orbital variations.  The
counterpart to source CX1 showed a $\sim 1.3$~mag total range of
variability, which is consistent with a dwarf nova eruption.  The ABs
showed substantially less evidence of variability, with only the
counterparts to CX8, CX19, and CX27 falling significantly above the
$\sigma(\ha)$-\ha\ relation for all stars.  Of these three, the two
counterparts to CX8 appear to be a foreground objects that are not
associated with the cluster.

Our determination of the cluster center agreed well with previous
recent determinations.  We found that while the radial profile of a
MSTO star group can be acceptably fitted with a cored power law out to
the half-mass radius, this model falls systematically below the
profile in the core and above it at intermediate radii.  In addition,
the slope for this cored-power-law fit does not agree with that of an
analytic King model.  Thus, we conclude that NGC~6752 does not show a
normal King model profile, in agreement with the previous findings of
\citet{Djorgovski86}, \citet{Ferraro03a}, and \citet{Thomson12}.  The
profile is better fit, out to a projected radius of 0.5~pc, by a cored
power law with a slope that is consistent with the typical value for a
post-core-collapse profile.  This supports the conclusion that
NGC~6752 is in a post-collapse state.

We compared the radial distributions of several different stellar
groups to that of a MSTO sample. We found that radial distributions of
all of the \chandra\ sources, the bright CVs, and the BSs show
a strongly significant central concentration relative to that of the
MSTO group.  We performed fits of a cored-power-law model to the
individual groups, in order to estimate the characteristic individual
stellar mass for each group.  We found that the \chandra\ sources, the
bright CVs, and the BSs all have a characteristic mass that
significantly exceeds the MSTO mass.  In the case of the bright CVs,
the characteristic mass of $1.6 \pm 0.3\,\msun$ is the similar to what
we found for NGC~6397 \citep{Cohn10}.

We found that the bright CVs are more centrally concentrated than the
faint CVs, consistent with a picture in which bright CVs represent a
population that has been recently formed by dynamical interactions
near the cluster center. The faint CVs then would represent an evolved
population that has been scattered out of the cluster core over
time. We find that, like the core-collapsed cluster NGC 7099, NGC 6752
is deficient in X-ray sources relative to the mean relation between
X-ray source population size and encounter rate. This supports the
suggestion that core-collapsed clusters underproduce X-ray binaries,
implying that binary destruction/ejection is more vigorous in
core-collapsed clusters.

\acknowledgements{We thank N.~Ivanova for providing unpublished
  details on her simulations of CVs in globular clusters. This work is
  supported by NASA grant HST-GO-12254.02-A to Indiana
  University. Phyllis Lugger and Haldan Cohn acknowledge the
  hospitality of the Department of Astronomy and Astrophysics at the
  University of California Santa Cruz, where part of this work was
  carried out.  Craig Heinke is supported by a NSERC Discovery Grant
  and a NSERC Discovery Accelerator Supplement Award.}

\onecolumngrid
\vspace*{12pt}
\software{IRAF,PyRaF,SExtractor,wavdetect,pwdetect,XSPEC} 

\twocolumngrid

\end{document}